\DocumentMetadata{pdfstandard=A-2b} 
\documentclass[pre,superscriptaddress,notitlepage,nofootinbib, twocolumn,dvipsnames]{revtex4-2}
\pdfoutput=1
\usepackage[left=0.6 in, right=0.6 in, top=0.7 in, bottom = 0.7 in]{geometry}
\usepackage{hyperref}
\usepackage{graphicx}
\usepackage{amsmath,amssymb,mathtools,bm, physics, ifthen, xparse}
\usepackage{pdfx}

\hypersetup{
    colorlinks=true,
    linkcolor=blue,
    filecolor=magenta,      	
    urlcolor=cyan,
}

\newcommand{\expv}[1]{\left\langle #1 \right\rangle}
\newcommand{\Fex}{\mathcal{F}^{\rm ex}}
\newcommand{\Fid}{\mathcal{F}^{\rm id}}
\renewcommand{\r}{\vb r}
\newcommand{\p}{\vb p}

\RenewDocumentCommand{\fdv}{m m g}{%
    \IfNoValueTF{#3}{%
        \frac{\delta #1}{\delta #2}
    }{%
        \frac{\delta^2 #1}{\delta #2 \delta #3}
    }%
}

\begin{document}

\title{Quantum Ornstein-Zernike Theory for Two-Temperature Two-Component Plasmas}

\author{Zachary A.~Johnson}
  \email{john8248@msu.edu}
  \affiliation{Computational Mathematics, Science and Engineering, Michigan State University, East Lansing, Michigan 48824, USA}%
\author{Nathaniel R.~Shaffer}
  \affiliation{Laboratory for Laser Energetics, University of Rochester, 250 East River Road, Rochester, New York 14623, USA}%
\author{Michael S.~Murillo}%
  \affiliation{Computational Mathematics, Science and Engineering, Michigan State University, East Lansing, Michigan 48824, USA}

\date{\today}

\begin{abstract}
Laboratory plasma production almost always preferentially heats either the ions or electrons, leading to a two-temperature state. In this state, density functional theory molecular dynamic simulation is the state of the art for modeling bulk material properties. We construct a statistical mechanics model for the two temperature limit that is theoretically consistent with the molecular dynamics method. We proceed to derive the electron-ion multi-temperature quantum Ornstein-Zernike equations for the first time. This allows the construction of a two-temperature two-component plasma model using the average atom from which we can compute bulk material properties at a fraction of the computation time of the two-temperature density functional theory simulation. The accuracy of the model is benchmarked against ion pair correlation and self-diffusion results from ab initio simulation. We proceed to compute the viscosity and ion thermal conductivity as a function of both ion and electron temperature.
\end{abstract}

\maketitle


\section{Introduction}
\label{sec:Intro}
Energy transport and equation of state are crucial physical inputs to accurate modeling of astrophysical environments and terrestrial experiments, including fusion energy efforts. The highly transient nature of inertial fusion energy and laser heated plasma experiments produces plasmas where the electrons and ion species are almost always initially at different temperatures. Hydrodynamic modeling typically takes into account temperature separation \cite{Marinak2001, Fryxell2000, Orban2013} but the hydrodynamic equations must be closed by calling multi-temperature models for state properties and transport properties.

The current state-of-the-art tool for dense plasma study is density functional theory molecular dynamics (DFT-MD), but typically under the assumption of an equilibrium plasma. However, in principle, it is also capable of modeling two-temperature electron-ion plasmas, and provides a crucial high-fidelity tool for the study of these systems, having been successfully applied to the study of two-temperature plasmas \cite{Fletcher2022,PhysRevLett.126.065001, PhysRevLett.95.085002, PhysRevB.77.075133}. These many-ion calculations are fundamentally limited in scope due to computational expense; this limits direct comparison with experiments in cases where the parameter space is large. In the pure equilibrium limit, this niche is filled by the much more rapid average atom (AA) models which provide an all electron but single ion approximation of the DFT-MD calculation. Many iterations of these models exist \cite{Wilson2006, Starrett2019, Starrett2014, Chihara2015,PhysRevResearch.3.023026, Hansen2023, callow2022atomec, PhysRevE.83.026403}, but any that include ion-ion correlations must appeal to the Ornstein-Zernike (OZ) equation and a closure such as the hypernetted-chain (HNC) approximation. These equations are necessary for modeling ionic equation of state contributions, the ion acoustic spectral feature, as well as for electronic and ionic transport coefficients\cite{Grabowski2020, Stanek2024}.   

Previous two-temperature generalizations of the average atom require an ansatz \cite{DW2T-2012,PhysRevE.58.3705} for how the electron and ion temperatures enter into the neutral pseudo atom (NPA) model or require constructing ion-ion potentials by other means such as through the Gordon-Kim approach \cite{gordon1972theory, HOU201721}. In this paper, motivated by \cite{BoerckerMore1986}, we assume an adiabatic electron picture and develop the fully strongly coupled and quantum generalization of the two-temperature quantum OZ and HNC equations which reduce to the standard forms in the equal temperature limit. This provides a rigorous theoretical framework for the generalization of AA models with ions and electrons to the two-temperature limit in a thermodynamically consistent way. In the limit where we replace quantum HNC electron-ion correlations with approximate closures based on the AA model, our model reduces to the two-temperature NPA model of \cite{DW2T-2012}, a main result of this paper.

It has been shown that the equilibrium NPA result is in good agreement with equilibrium density functional theory computations \cite{stanek2021efficacy}. It has additionally been used to calculate non-equilibrium equation of state\cite{harbour2017equation}, acoustic modes \cite{harbour2018ion}, and electronic transport \cite{dharma2017isochoric}, but has not been similarly validated against explicit two-temperature DFT-MD calculations as in the equilibrium case. We take the AA model in \cite{Starrett2014}, and show how it and other AA models can be generalized to the two-temperature limit as in the NPA model, to generate a two-temperature two-component plasma (2TTCP) model. We then compare it against radial distribution functions and self-diffusion with two-temperature \textit{ab initio} DFT-MD simulation, finding excellent agreement. We further compute the ion thermal conductivity and ion viscosity from molecular dynamics, comparing against the results of \cite{HOU201721}. We also determine the reliability of on-the-fly ionic transport coefficient models in the two-temperature limit \cite{PhysRevE.93.043203, MURILLO200849, Johnson2024}. The validation of the 2TTCP approach as a generalization of equilibrium AA models against DFT-MD, and the prediction of non-equilibrium ionic transport coefficients are the other main results of this paper.

In section~\ref{sec:Theory} we introduce our multi-temperature partition function and the accompanying thermodynamic framework. We then derive the two temperature quantum Ornstein-Zernike (QOZ) and hypernetted-chain (HNC) equations. In section~\ref{sec:2TTCP AA} we use these ideas to create a two-temperature, two-component plasma model based on the average atom. We then compute ionic transport coefficients, and radial distribution functions, comparing results with the literature. 


\section{Multiple Temperature Statistical Mechanics}
\label{sec:Theory}
Materials with multiple temperatures are by definition out of local thermal equilibrium (LTE). However, in situations where the equilibration timescales are hierarchical, the dynamics at a given timescale are approximately equilibrated and well described by a partition function adiabatically evolving in time. In the case of an electron-ion plasma, this hierarchy is $\tau_{ee}<\tau_{II} < \tau_{eI}$ where $\tau_{ij}$ is the thermodynamic relaxation time between species $i$, $j$. We denote electrons and ions by $e$, $I$, respectively. This leads to an adiabatic electron description used for example in the DFT-MD codes\cite{Giannozzi_2009, PhysRevB.54.11169, Gonze2020}. The electrons in this picture relax in the fixed ion potential, with the Hamiltonian (in atomic units, $\hbar=m_e=e=k_B=1$)
\begin{align}
    \hat{H}_{e} &= \hat{H}_{ee} + \hat{H}_{eI} \\
    &= \frac{1}{2} \sum_i \hat{p_i}^2 +  \frac{1}{2} \sum_{i\neq j} \frac{1}{| \hat{\r}_i - \hat{\r}_j|} + \sum_i \hat{V}^{eI}(\{\r_I \},\hat{\r}_i) \nonumber\\
    &\ +\sum_i \hat{V}_e^{\rm ext}(\hat{\r}_i)), \nonumber
\end{align}
where $\hat{V}_e^{\rm ext}$ is the external potential on the electrons, and $V^{eI}$ is the electron-ion potential from ions fixed at coordinates $\r_I$ which are the classical degrees of freedom alongside the ion momenta, which do not enter into the electron dynamics in the adiabatic approximation. As such, to simplify the math we will refer to the quantity,
\begin{align}
\label{eq:ψe}
    \hat{\psi_e}(\hat{\r}_i) = \mu_e  - \hat{V}^{eI}_e(\{\r_I\}, \hat{\r}_i) - \hat{V}^{\rm ext}_e(\hat{\r}_i),
\end{align}
as the external potential on the electrons though it includes the electron chemical potential $\mu_e$. This specifies the partition function for the electrons,
\begin{align}
\label{eq:electron_partition}
    \Xi_e &= \exp{-\beta_e \Omega_e} \nonumber \\
         &= \sum_{N_e} {\rm Tr } e^{-\beta_e ( \hat{H}_{ee} -\sum_i \hat{\psi_e}(\hat{\r}_i) )}.
\end{align}
with electron inverse temperature $\beta_e$, number of electrons $N_e$, and grand potential $\Omega_e$. 
This fixes the electronic physics, but the more nontrivial step is determining the statistical mechanics of the ions in the presence of these electrons. The ion-ion Hamiltonian itself is 
\begin{align}
    H_{II} = \sum_{i}\frac{{\vb p}_i^2}{2 m_I} + \sum_{i\neq j} \frac{1}{2} \frac{Z_I^2}{|\r_i - \r_j |},
\end{align}
Since the electrons minimize the grand energy for constant temperature and chemical potential, in the adiabatic limit it is the grand energy that affects the forces and energetics of the ion subsystem. The electronic grand potential, an implicit function of the electron temperature, then acts as a new energy term in the ion partition function, weighted by the inverse ion temperature $\beta_I$. The partition function in the grand canonical ensemble is then,
\begin{align}
\label{eq:BM_partition_function}
    \Xi = \sum_{N_I} \int \prod^{N_I}_i \frac{ d^3\r_i d^3\p_i}{(2\pi\hbar)^{3}} e^{-\beta_I ( H_{II} + \Omega_{e}(\{\vb r_i\};\beta_e)  - \sum_i \psi_I(\r_i)    )},
\end{align}
where $N_I$ is the number of ions in a given configuration and the sum is over all positive integers, and we refer to $\psi_I(\r_i) = \mu_I - V_I^{\rm ext}(\r_i)$  as the external potential on the ions as in Eq.~\eqref{eq:ψe} for the electrons, with the key difference that $\psi_I$ only includes true external potentials from e.g. external electric fields, whereas $\psi_e$ includes the ions themselves as a source as well. This idea is essentially the ansatz of \cite{BoerckerMore1986} where it was applied to study classical, weakly interacting particles. In App.~\ref{app:thermo derivation} we present a statistical derivation of this partition function and increase the connection with DFT-MD simulations.  
The grand energy for the full system is 
\begin{align}
\Omega = - T_I \ln \Xi.\label{eq:Ω_statistical}
\end{align}
From the grand potential we also have a free energy, $F$ corresponding to  a generalization of the Helmholtz free energy to two-temperature systems. These are related via
\begin{align}
\label{eq:thermo_Ω}
\Omega &= F - \sum^N_j \mu_j \int d^3\r n_j,
\end{align}
where $\mu_j$ is the chemical potential and $n_j$ is the number density of species $j \in \{e,I\}$. Each species will have its own external potential $V^{\rm ext}_j$. A more useful object here is the intrinsic free energy, which in an abuse of notation, we will simply refer to as the free energy, 
\begin{align}
\mathcal{F} &= F - \sum_j^N \int d^3\r V^{\rm ext}_j n_j .
\end{align}
This is the Legendre transform of the grand energy with respect to $\psi_j$,
\begin{align}
    \Omega = \mathcal{F} - \sum_j^N  \int d^3 \r n_j \psi_j. \label{eq:Legendre}
\end{align}
We must show that this thermodynamic relationship is consistent with the statistical definition in Eq.~\eqref{eq:Ω_statistical}. We can see this by taking functional derivatives of the partition function, generating
\begin{align}
\label{eq:ni}
    \fdv{\Omega}{\psi_I(\r)} 
                &=  - \langle  \sum_k \delta^3(\r - \r_k)\rangle_I, \nonumber \\
                &=- \langle \hat{n_I} \rangle_I,  \nonumber\\
                &\equiv -n_I(\r).
\end{align}
where the density operator is defined implicitly in the second line, and we define the ion configuration average as 
\begin{align}
    \langle \mathcal{O} \rangle_I = \frac{1}{\Xi} \sum_{N_I} \int \prod^{N_I}_i \frac{ d^3\r_i d^3\p_i}{(2\pi\hbar)^{3}} e^{-\beta_I ( H_{II} + \Omega_{e}  - \sum_i \psi_I(\r_i)    )} \mathcal{O}.
\end{align}
This shows the consistency of the ionic component. For the electrons, we have
\begin{align}
\label{eq:ne}
    \fdv{\Omega}{\psi_e(\r)} 
                &=  \Big\langle \fdv{\Omega_e[\{ \r_I\}  ]}{\psi_e(\r)}   \Big\rangle_I,\nonumber \\
                &=  \Big\langle  - \langle \hat{n}_e \rangle_e(\{ \r_I \}) \Big\rangle_I, \nonumber\\
                &=  - \langle n_e(\{ \r_I\})   \rangle_I, \nonumber \\ 
                &\equiv -n_e(\r).
\end{align}
The electronic averaging is done for a specific ion configuration via,
\begin{align}
     \langle \mathcal{O} \rangle_e (\{ \r_I \})= \frac{1}{\Xi_e} \sum_{N_e} {\rm Tr } e^{-\beta_e ( \hat{H}_{ee} -\sum_i \hat{\psi_e}_i  )} \mathcal{O}.
\end{align}
 Note we have to keep track of two electron densities, the specific ion configuration density $n_e(\r;\{ \r_I\})$, and the ion-averaged quantity, $n_e(\r)$. 
 
 To make this theory useful, we must derive relations between equilibrium correlation functions through analogues of the OZ and HNC equations described in the next few subsections. 

\subsection{Two-Temperature Quantum Ornstein-Zernike Relations}

Fundamental to the Ornstein-Zernike relations is the separation between the ideal, or non-interacting, free energy components and the rest, or excess as\cite{HansenMcdonald}
\begin{align}
    \mathcal{F} = \Fid + \Fex.
\end{align}
This idea originates in classical statistical mechanics since this is paramount to the separation of kinetic and potential degrees of freedom, yet it extends quite naturally to the framework of Kohn-Sham density functional theory wherein a non-interacting reference system is used. 

We proceed by defining the response function and the polarization potential, respectively, as
\begin{align}
    \fdv{\Omega}{\psi_i(\r)}{\psi_j(\r')} = -\fdv{n_i(\r)}{\psi_j(\r')} &\equiv  \chi_{ij}(\r,\r'), \label{eq:χ_def}\\
    \frac{\delta^2 \Fex[\{n\}]}{ \delta n_{i}(\r) \delta n_{j}(\r')} = \fdv{\psi_i(\r)}{n_j(\r')} + {\chi^0_{ij}}^{-1}(\r,\r')  &\equiv U_{ij}(\r,\r' )\label{eq:U_def}.
\end{align}
In fully equilibrated systems the polarization potential $U_{ij}$, is related to the direct correlation function as $c_{ij}=-\beta U_{ij}$. Working with the polarization potential instead allows us to remain agnostic about the nature of $\beta$ when $T_i \neq T_j$.
Constructing the Ornstein-Zernike equation begins with the definition of the functional inverse, 
\begin{align}
\label{eq:functional_χ_inverse}
    \sum_{k}\int  \chi_{ik}(\r,\r'')  \chi^{-1}_{ki}(\r'',\r')  d\r''= \delta_{ij}\delta(\r-\r').
\end{align}
For homogeneous systems, we write this in k-space and as a matrix equation over species space, giving us the Dyson equation for the response function,
\begin{align}
\label{eq:Dyson}
    \tilde{\pmb \chi}(k) = (\tilde{\pmb \chi^0}^{-1}(k) - \tilde{\pmb U}(k) )^{-1}.
\end{align}
Given the theoretical importance of electron-ion plasma, it is instructive to consider the specific case of a two species plasma, for which we have, 
\begin{align}
\label{eq:χOZ_2species}
    \tilde{\pmb \chi}(k)= \frac{1}{D}
    \begin{pmatrix}
           {\chi^0}_{11}( 1  - {\chi^0}_{22} U_{22})  & {\chi^0}_{11}{\chi^0}_{22} U_{12} \\
             {\chi^0}_{11}{\chi^0}_{22} U_{12} & {\chi^0}_{22} ( 1 - {\chi^0}_{11} U_{11})
    \end{pmatrix} .
\end{align}
where the denominator is,
\begin{align}
    D =  ( 1 - {\chi^0}_{22} U_{22})(1 - {\chi^0}_{11} U_{11}) - {\chi^0}_{11}{\chi^0}_{22} U_{12}^2 .
\end{align}
This reduces to the equilibrium answer\cite{PhysRevE.101.013208} when $U_{ij}\to -\beta_i c_{ij}.$ 

Turning this Dyson equation into the useful form of the OZ equations requires deriving relations between the response function and particle-particle correlation functions. 
The second functional derivative generates the density-density correlation functions through 
\begin{align}
        \fdv{\Omega}{\psi_i(\r') }{\psi_j(\r)} &= -\frac{1}{\beta_I \Xi}\fdv{\Xi}{\psi_i(\r') } {\psi_j(\r)} + \frac{1}{\beta_I}\fdv{\ln \Xi}{\psi_i(\r') }\fdv{\ln \Xi}{\psi_j(\r') }.
\end{align}
The ion-ion term is\cite{HansenMcdonald}, 
\begin{align}
\label{eq:χii}
    \fdv{\Omega}{\psi_I(\r') }{\psi_I(\r)} &= -\beta_I \langle \hat{n_I}(\r) \hat{n_I}(\r')\rangle_I -  \beta_I n_I(\r)n_I(\r'), \nonumber \\
        &\equiv -\beta_I H_{II}(\r,\r'),
\end{align}
Where $H_{ij}$ describes fluctuations in the density-density correlation about the background, 
\begin{align}
H_{II}(\r,\r') = \big\langle  \left(\hat{n}_I(\r) - \expv{\hat{n}_I(\r)}_I \right)\left(\hat{n}_I(\r') - \expv{\hat{n}_I(\r')}_I \right) \big\rangle_I.     
\end{align}
The ion-electron term is likewise
\begin{align}
\label{eq:χie}
    \fdv{\Omega}{\psi_i(\r') }{\psi_e(\r)} 
    &= -\beta_I \Big\langle \hat{n}_I(\r')\fdv{\Omega_e[\{ \r_I\}  ]}{\psi_e(\r)} \Big \rangle_I +  \beta_I n_I(\r')n_e(\r), \nonumber \\
    &\equiv -\beta_I H_{Ie}(\r,\r'),
\end{align}
where one can check explicitly that this functional derivative is symmetric upon interchange of $\psi_I$ and $\psi_e$, and 
\begin{align}
H_{Ie}(\r,\r') = \big\langle  \left(\hat{n}_I(\r) - \expv{\hat{n}_I(\r)}_I \right)\left(n_e(\{ \r_I\})  - \expv{n_e(\{ \r_I\}) }_I \right) \big\rangle_I.     
\end{align}
The electron-electron term is additionally complicated by the need for a chain rule operation on the exponential of the electron grand potential, resulting in  
\begin{align}
\label{eq:δ2Ωδee}
        \fdv{\Omega}{\psi_e(\r') }{\psi_e(\r)} 
        &=  \Big\langle \fdv{\Omega_e[\{ \r_I\}  ]}{\psi_e(\r)}{\psi_e(\r')} - \beta_i  \fdv{\Omega_e[\{ \r_I\}  ]}{\psi_e(\r)} \fdv{\Omega_e[\{ \r_I\}  ]}{\psi_e(\r')}  \Big\rangle_I \nonumber \\
        &+  \beta_I n_e(\r')n_e(\r).
\end{align}
This first term in Eq.~\eqref{eq:δ2Ωδee} is the density-density fluctuation term in a specific fixed ion configuration, 
\begin{align}
    \fdv{\Omega_e[\{ \r_I\}  ]}{\psi_e(\r)}{\psi_e(\r')} 
    = -\beta_e H_{ee}(\r, \r'; \{ \r_I \}).
\end{align}
which is,
\begin{align}
\label{eq:Hee_notaveraged}
H_{ee}(\r,\r'; \{\r_I\}) = \big\langle \hat{n}_e(\r) \hat{n}_e(\r') \big\rangle_e - n_e(\r;\{\r_I\})n_e(\r';\{\r_I\}).     
\end{align}
Inserting this into Eq.~\eqref{eq:δ2Ωδee} gives
\begin{align}
     \fdv{\Omega}{\psi_e(\r') }{\psi_e(\r)} &= -\beta_e \langle \hat{n}_e(\r) \hat{n}_e(\r') \big\rangle_{e, I} \nonumber \\
     &+ (\beta_e-\beta_I) \langle n_e(\r;\{\r_I\})n_e(\r';\{\r_I\})\rangle_{I}\nonumber \\ &+  \beta_I n_e(\r')n_e(\r).
\end{align}
The second term on the right hand side is a newly relevant correlation function, being the ion-averaged correlation between one-body electron densities. It is a perfectly well-defined function even in the equal temperature limit, but never enters. 

To accommodate this new correlation structure, we define another electron-electron density-density fluctuation function,
\begin{align}
\label{eq:Hee_averaged}
H_{ee}(\r,\r') = \big\langle \hat{n}_e(\r) \hat{n}_e(\r') \big\rangle_e - n_e(\r)n_e(\r'),
\end{align}
which is distinct from Eq.~\eqref{eq:Hee_notaveraged} in that the one-body densities have been averaged over ion configurations independently.

We have the difference, 
\begin{align}
\label{eq:ΔHee}
     \Delta H_{ee} (\r,\r')  &= H_{ee}(\r,\r') - \langle H_{ee}(\r,\r'; \{\r_I\})\rangle_I \nonumber \\ 
    &= \left\langle n_e(\r;\{\r_I\})n_e(\r';\{\r_I\}) \right\rangle_I  - n_e(\r)n_e(\r')  .
\end{align}
This implies 
\begin{align}
    \fdv{\Omega}{\psi_e(\r') }{\psi_e(\r)} 
    &= -\beta_e H_{ee} (\r,\r') + (\beta_e - \beta_I) \Delta H_{ee}(\r,\r')
\end{align}
This non-equilibrium correction to the electron-electron behaviour can be easily thought of in two extreme limits, the homogeneous electron gas (HEG), and the tight-binding limit. In the first case, the ions are spread homogeneously in space, so the one-body density is likewise homogeneous and the $\beta_I$ dependence cancels exactly. In the opposite limit of tight-binding electrons (TB), the electron number density is essentially composed of delta functions around the ions, and ion-ion correlations enter, as in

\begin{align}
    \Delta H_{ee} (\r,\r') &\xrightarrow[]{{\rm HEG }} 0 \nonumber \\
    \Delta H_{ee} (\r,\r')  &\xrightarrow[]{{\rm TB }}  Z^2 H_{II}(\r, \r') \nonumber.
\end{align}
We can also see that in the very cold ion limit, the ion-correlations become dominant.
These fluctuations are connected to the response function via Eq.~\eqref{eq:χ_def}, giving
\begin{subequations}
    \begin{align}
        \chi_{II}(\r,\r') &= -\beta_I H_{II}(\r,\r') \label{eq:χij_a}\\
        \chi_{Ie}(\r,\r') &= -\beta_I H_{Ie}(\r,\r') \label{eq:χij_b}\\
        \chi_{eI}(\r,\r') &= -\beta_I H_{Ie}(\r,\r') \label{eq:χij_c}\\
        \chi_{ee}(\r,\r') &= -\beta_e H_{ee}(\r,\r') + (\beta_e - \beta_I) \Delta H_{ee}(\r,\r')\label{eq:χij_d}.
    \end{align}
\end{subequations}
We now construct a form of the Ornstein-Zernike relations in terms of the polarization potential $U_{ij}$ in Eq.~\eqref{eq:U_def}, and the dimensionless pair correlation function,
\begin{align}
    h_{ij}(\r, \r') = \frac{H_{ij}(\r, \r')}{n_i(\r)n_j(\r')} - \frac{\delta_{ij} \delta(\r-\r')}{n_i(\r)},
\end{align}
as well as the dimensionless electron-electron deviation from the HEG limit 
\begin{align}
     \Delta h_{ee}(\r, \r') =  \frac{\Delta H_{ee}(\r, \r')}{n_e(\r)n_e(\r')}  - 1 
\end{align}
We now take the classical ion limit, $\chi^0_{II} = -\beta_I n_I$, and assume homogeneity. In Fourier space, from Eq.~\eqref{eq:χOZ_2species} we get,
\begin{subequations}
\label{eq:final_OZ_form1}
    \begin{align}
        \tilde{h}_{II}(k) &= -\frac{1}{D} \beta_i (U_{II}  - {\chi^0}_{ee}( U_{II} U_{ee} - U_{eI}^2 ) )  \\
        \tilde{h}_{Ie}(k) &= \frac{1}{D} n_e^{-1}{\chi^0}_{ee} U_{Ie}\\
        \tilde{h}_{eI}(k) &= \frac{1}{D} n_e^{-1} {\chi^0}_{ee} U_{eI} \\
        \tilde{h}_{ee}(k) &= (1-\beta_I/\beta_e) \widetilde{\Delta h}_{ee} - n_e^{-1} (1 + n_I \beta_I \tilde{U}_{II})(1 + \frac{\chi^0_{ee}}{\beta_e n_e}) \nonumber \\ &  +\frac{1}{D} n_e^{-1} {\chi^0}_{ee} (U_{ee}+ n_I \beta_I (U_{II}U_{ee} - U_{eI}^2)) .
    \end{align}
\end{subequations}
These are the two-temperature generalizations of the quantum Ornstein-Zernike relation and is a major result of this paper. To see another way in which they can be written, consults App.~\ref{app:other_QOZ}. One can see that in the equilibrium limit $\beta_I=\beta_e$ the new correlation term $\widetilde{\Delta h}_{ee}$ drops  out, and taking the classical limit for the electrons makes it symmetric with the ions. One can then recover the equilibrium OZ relations via $U_{ij}\to \beta c_{ij}$.

\subsection{Hypernetted-Chain for Two-Temperature Systems}
We have connected the response function to spatial correlation functions, but as of yet we require a closure that connects the polarization potential to these functions. In a classical equilibrium system, this is typically in the form of the hypernetted chain closure, with or without corrections\cite{rosenfeld1979theory, PhysRevA.46.1051, PhysRevE.61.2129}. In many two-temperature papers \cite{BoerckerMore1986, Shaffer2017, SVT1989} this closure is commonly referenced, but a derivation has never been given to our knowledge, nor is it obvious what the multi-temperature generalization of the direct correlation function $c_{ij}$ is supposed to be. It would seem particularly troubling since the canonical form of the HNC approximation, 
\begin{align}
    h_{ij}+1 = e^{-\beta u_{ij} + h_{ij} - c_{ij}} \label{eq:normal HNC}
\end{align}
implicitly incorporates the Ornstein-Zernike relations\cite{HansenMcdonald}, which were modified in \cite{SVT1989, Shaffer2017}.

In this section, we derive the multi-temperature generalization of Eq.~\eqref{eq:normal HNC}. In principal, via thermodynamic integration over density variations one can get an exact expression relating the direct correlation and pair distribution functions\cite{HansenMcdonald}, but it is simpler to approximate the polarization potential in the homogeneous limit, which defines the HNC approximation. To do this we vary the density in a second order Taylor expansion of the free energy around a uniform background,
\begin{align}
    n_i(\r) = {n_0}_i + \Delta n_i(\r).
\end{align}
In terms of the polarization potential Eq.~\eqref{eq:U_def}, we have 
\begin{align}
    \Fex[\{n\}]=& \Fex[\{n_0\}]  + \sum_i \mu^{\rm ex} \int d\r \Delta n_i(\r) \nonumber \\ 
    & + \frac{1}{2} \sum_{ij} \int d\r d\r' U_{ij}(\r,\r') \Delta n_i(\r) \Delta n_j(\r') + \sum_{ij} \mathcal{B}_{ij}
\end{align}
Where we use the excess chemical potential $\mu^{\rm ex} = \fdv{\Fex[{n_0}]}{n_i} = \psi - \fdv{\mathcal{F}^{\rm id}[{n_0}]}{n_i} = \mu - \fdv{\mathcal{F}^{\rm id}[{n_0}]}{n_i}$, and the functional $\mathcal{B}_{ij}$ encodes the difference between the homogeneous polarization potential and the exact heterogeneous polarization potential one could obtain from a coupling constant integration.  
Here we can see the advantage of using the polarization potentials rather than the direct correlation functions as we as of yet do not have to specify any temperatures.

Inserting this expression into grand energy Eq.~\eqref{eq:Legendre}, the Euler equation we get is,
\begin{align}
    0 =& \fdv{\Omega}{n_i(r)} \\
      =& \fdv{\Fid[n_i]}{n_i(r)} + \mu^{\rm ex} - \psi_i(\r)\nonumber \\  &+  \sum_j \int d\r' U_{ij}(\r,\r') \Delta n_j(\r') + B_{i},
\end{align}
where we have implicitly defined the bridge function $B_{i} = \sum_j \fdv{\mathcal{B}_{ij}}{n_i(r)}$, which we define with only one subscript since we note that it depends explicitly on all species. The discussion thus far is general to quantum and classical systems, but now the electrons and ions must be separately handled. In the classical limit, for the ions, we have 
\begin{align}
    0 &= T_i \ln \left( n_i(\r)/{n_0}_i \right) + \sum_j \int d\r' U_{ij}(\r,\r') \Delta n_j(\r') +\nonumber\\ &V^{\rm ext}(\r) + \sum_j \fdv{B_{ij}}{n_i(r)}.
\end{align}
For the electrons, we instead take the Kohn-Sham picture and create a non-interacting system in a modified effective potential\cite{Anta2000}
\begin{align}
    n_e\left(\r\right) &= n_e\left(\r| V^{\rm eff\ ext}\right) \\
    V^{\rm eff\ ext}(\r) &= V^{\rm ext}_i(\r) + \sum_j \int d\r' U_{ij}(\r,\r') \Delta n_j(\r'),
\end{align}
with a chemical potential $\mu^0_i = \mu_i + \mu_i^{\rm ex}$. 
These densities are connected with correlation functions through Percus's trick\cite{HansenMcdonald}
\begin{align}
    n_i(r; u_{ij}) = {n_0}_i  g_{ij}^{(2)}(r),
\end{align}
where the external potential is identified as the inter-particle potential $u_{ij}$---the density around a particle of species $j$ fixed at the origin is proportional to the homogeneous two particle correlation function. It is only possible to fix classical particles at the origin in this way, but the number density can be for any species. This allows us to obtain $g_{eI}$, $g_{II}$, but not $g_{ee}$, whose closure is unknown.\footnote{We note for the reader interested in purely classical systems, that Percus's trick does not work for the lighter species in this two-temperature scenario, as instead the usual trick \cite{HansenMcdonald} leads to non-trivial and likely useless relations between fractional powers of correlation functions. This is likely because fixing the position of a light species particle is fundamentally incompatible with the assumption of allowing it to adiabatically move in relation to the heavier species.} 

With the ion fixed at the origin, we obtain
\begin{subequations}
\label{eq:HNC_both}
    \begin{align}
        g_{II}(\r) &= \exp\Big[ -\beta_I u_{II}(\r) - \sum_{j={e,I}} n_j \int d\r' \beta_I U_{Ij}(\r,\r')  h_{jI}(\r') ) \nonumber \\ 
        &+ B_{I} \Big], \label{eq:HNC_ii}\\
        g_{eI}(\r) &= {n_0}_e^{-1}\ n_e\Big(\r| u_{eI}(\r) + \sum_{j={e,I}} n_j \int d\r' U_{ej}(\r,\r') h_{jI}(\r') \nonumber \\ 
        & B_e \Big) \label{eq:HNC_ie}.
    \end{align}    
\end{subequations}
The ion-ion equation can be simplified even further through a rewritten form of the QOZ equations Eq.~\eqref{eq:ion-ion_OZ_traditional}, to obtain
\begin{align}
\label{eq:final_gii_HNC}
       g_{II}(\r) &= \exp\left[ -\beta_I u_{II}(\r)  + h_{II} + \beta_I U_{II} + B_I \right]. 
\end{align}
This is remarkably the exact standard form of the HNC approximation if one replaces $\beta_I U_{II}$ with $-c_{II}$ and sets $B_I=0$, which is likely why models such as \cite{SVT1989} are so successful in the classical limit.  
The ion-ion HNC approximation in Eq.~\eqref{eq:final_gii_HNC} is the only HNC approximation we make. Incorporating the HNC equations for the electron-ion term would create a so-called QHNC model as in \cite{Chihara1978,Anta2000}. Fundamentally this model offers a complete picture with precisely described correlation structure in the two-temperature limit. However, it requires an expensive self-consistent calculation, and secondly the Taylor expansion in density fluctuations used is a quite poor approximation for the bound states, meaning the unknown $B_e$ bridge function could be quite large. Instead we opt for a different electron-ion closure\cite{Starrett2014} which does not require knowing the bridge functions and can be done quite accurately without a full self-consistent electron-ion iteration loop. 


\section{Two-Temperature Two-Component Plasma Model}
\label{sec:2TTCP AA}
The form of Eq.~\eqref{eq:final_gii_HNC}, excepting for a moment the bridge functions, involves only ion-ion correlations. This motivates an effective ion description which can be done exactly if the effective pair potential and polarization potential satisfy the relation
\begin{align}
    \tilde{u}_{II}^{\rm eff} - \tilde{U}_{II}^{\rm eff}&= \tilde{u}_{II} - \tilde{U}_{II}.
\end{align}
We also need an associated effective single particle OZ equation,
\begin{align}
    \tilde{h}_{II} = -\frac{\beta_I \tilde{U}_{II}^{\rm eff}}{1 + n_I \beta_I \tilde{U}_{II}^{\rm eff }}.
\end{align}
which is consistent with the full set of OZ equations Eqs.~\eqref{eq:final_OZ_form1} if we have the effective polarization potential 
\begin{align}
    \tilde{U}_{II}^{\rm eff} &= \tilde{U}_{II} + \frac{\tilde{\chi}^0_{ee} }{1-\chi^0_{ee} \tilde{U}_{ee}} \tilde{U}_{eI}^2, \nonumber \\
    &= \tilde{U}_{II} + n^{\rm scr} \tilde{U}_{eI} \label{eq:Ueff_from_nscr},
\end{align}
and the corresponding effective pair potential 
\begin{align}
        \tilde{u}_{II}^{\rm eff} &= \tilde{u}_{II} + n^{\rm scr} \tilde{U}_{eI},\label{eq:uII_eff}
\end{align}
where we have defined a screening potential\cite{Starrett2014,Chihara_1985}, 
\begin{align}
     n^{\rm scr} = \frac{\tilde{\chi}^0_{ee} }{1-\chi^0_{ee} \tilde{U}_{ee}} \tilde{U}_{eI}.
\end{align}
The relation between this screening density and a Kohn-Sham AA calculation is not precisely known, and thus constructing $n^{\rm scr}$ is where the ambiguity arises. We use the model in \cite{Starrett2014} which takes the difference between the free electrons of the AA calculation and an identical calculation with no core ion. We work additionally in a single-shot picture where the average atom computation is done with a fixed empty-core ion distribution function. This is shown to be a very good approximation \cite{Starrett2014}, and avoids the use of unknown and potentially large electron-ion and electron-electron bridge functions.  


\begin{figure}[t!]
    \centering
    \includegraphics[width=1\linewidth]{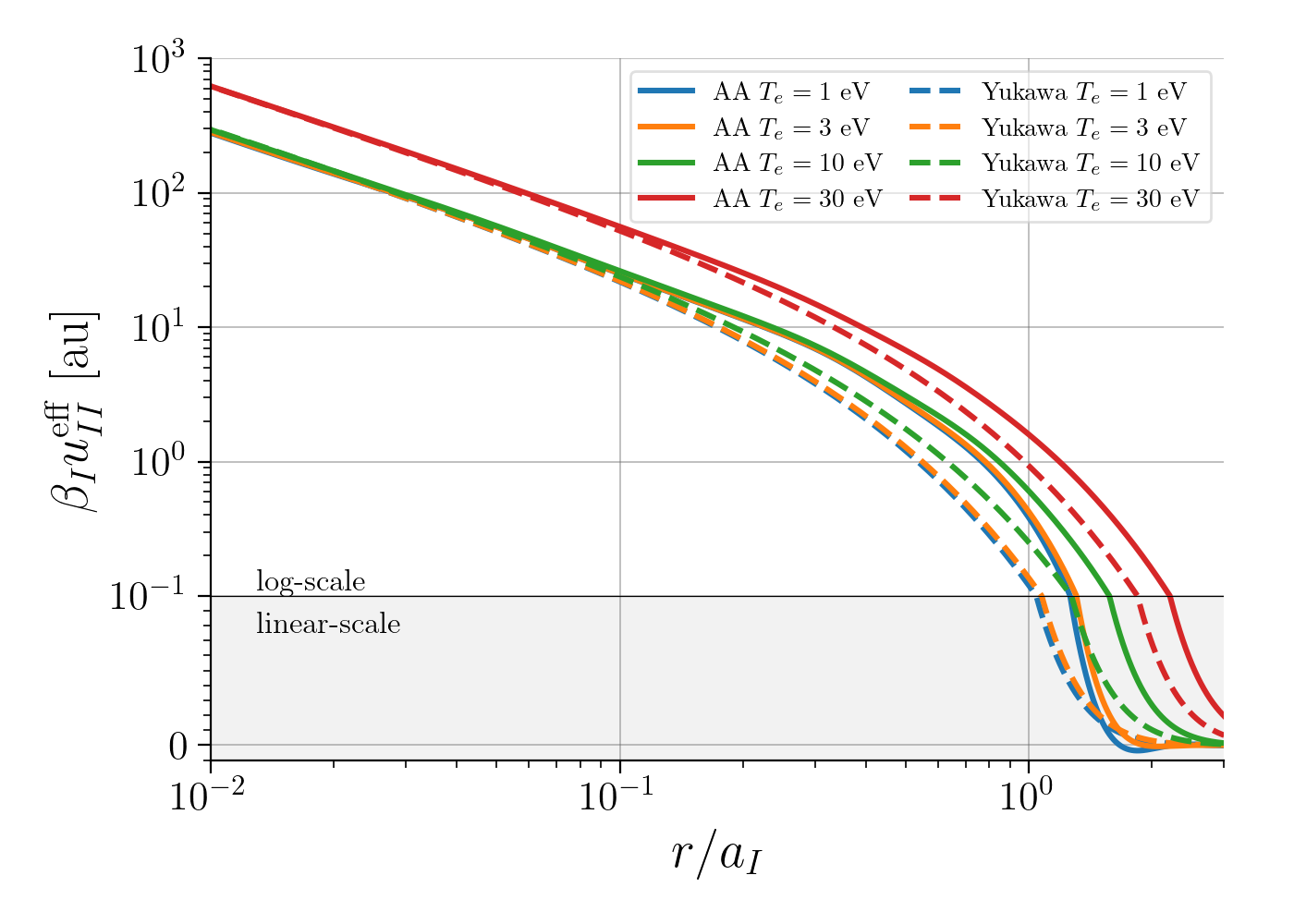}
    \caption{The AA produced pair potential \ref{eq:uII_eff} (solid lines), and Yukawa pair potential (dashed) with Thomas-Fermi screening using a $\langle Z \rangle$ from the AA model. Higher electron temperatures are the right most lines. The radial distance on the $x-axis$ is normalized to the ion sphere radius $a_I = (n_I 4\pi/3)^{-1/3}.$} 
    \label{fig:βu_comparison}
\end{figure}

\begin{figure*}[ht!]
    \centering
    \includegraphics[width=\textwidth]{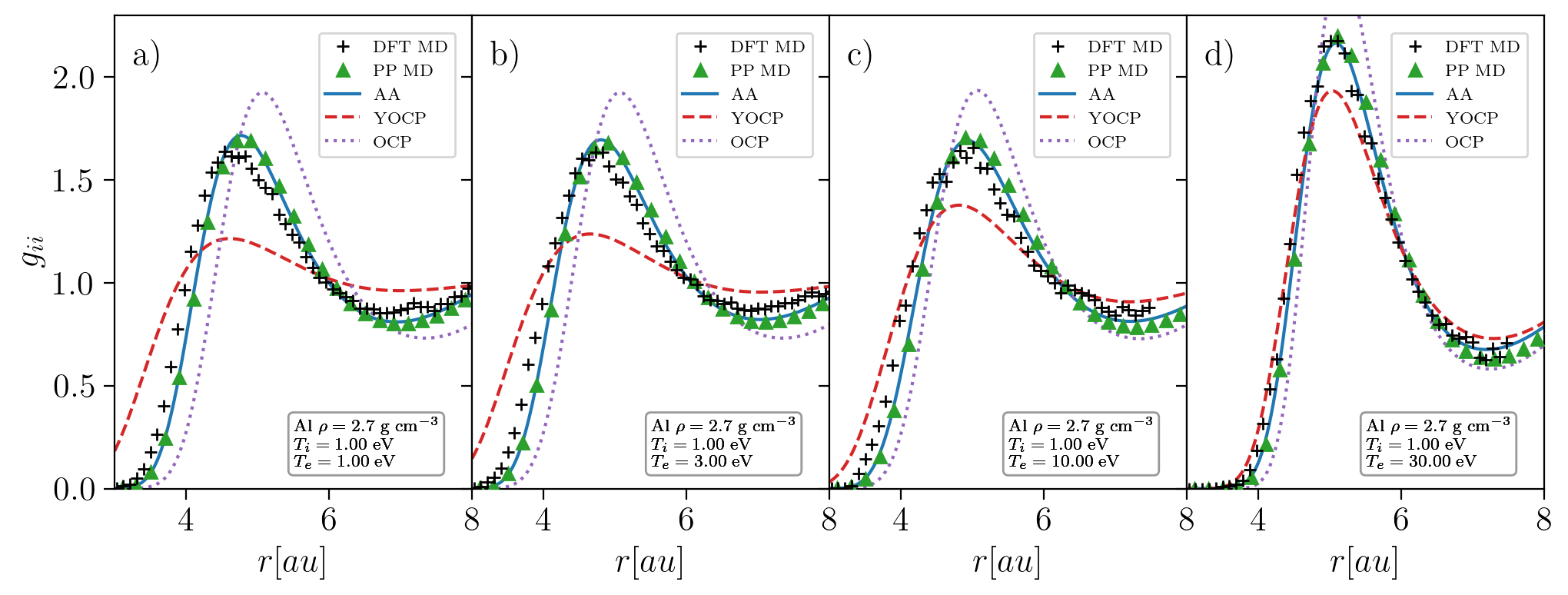}
    \caption{Ion-ion radial distribution function for Aluminum at solid density $2.7 {\rm g/cm}^3$ and fixed $T_I=1$ eV, with increasing electron temperature from left to right, {\bf a)} LTE: $T_e=1$ eV ($\langle Z \rangle=3.00 $), {\bf b)} $T_e=3$ eV ($\langle Z \rangle=3.00 $), {\bf c)} $T_e=10$ eV ($\langle Z \rangle=3.02 $), {\bf d)} $T_e=30$ eV ($\langle Z \rangle=4.35 $). We show DFT-MD (black plus), classical MD with our 2TTCP pair potential, and HNC with bridge function computations for the 2TTCP (blue solid), the YOCP (red dashed), and the OCP (purple dotted). }
    \label{fig:Al_fixed_Ti_structure}
\end{figure*}

Using any AA model, one can use Eq.~\eqref{eq:Ueff_from_nscr} to create the approximate electron-ion polarization potential to complete the effective potential. For example, this was done in the NPA picture by Dharma-Wardana in, for instance, \cite{DW2T-2012}.


\subsection{Plasma Structure}
\label{sec:plasma structure}
The electron and ion temperature are different levers on the structural and transport properties of dense plasmas. We investigate these effects through the 2TTCP model defined above, and compare to ab initio DFT-MD and on-the-fly transport models when possible. 

The effect of the electron temperature on the ions can be seen readily from the 2TTCP pair potential, Eq.~\eqref{eq:uII_eff}. In the case of a simple metal like aluminum with well defined bound and valence states, $T_e$ variations largely only impact ionic properties through ionization, which occurs at $\gtrsim 10$ eV. In Fig.~\ref{fig:βu_comparison} we can see both the effects of this ionization through the difference between the cooler (green, orange, blue) lines and the $T_e=30$ eV lines (red) as well as the importance of the non-linear screening in the 2TTCP (solid lines) versus the Yukawa one component plasma (YOCP) model (dashed lines). 

We benchmark the accuracy of our 2TTCP model with DFT-MD simulations using VASP \cite{PhysRevB.54.11169, PhysRevB.47.558, KRESSE199615} with an 11 electron projector augmented wave pseudopotential\cite{PhysRevB.59.1758}. The 11 electron pseudopotential  was necessary to capture ionization. The cutoff energy was 1 keV, and enough bands were included to ensure that the highest energy band had a Fermi-Dirac occupancy no greater than $10^{-6}$. The Brillouin zone was sampled at the Baldereschi mean-value point \cite{PhysRevB.7.5212}. For the $T_e=1, 3$ eV cases, we use 64 atoms, and for the $T_e=10, 30$ eV cases, we use 32 atoms. Ion dynamics in an NVT ensemble were simulated until well-converged radial distribution functions and mean-square displacements were obtained, between 844 and 2959 fs. 

We also run ion-ion pair potential MD using LAMMPS\cite{THOMPSON2022108171} and the 2TTCP pair potential (PPMD). We use $10,000$ atoms with timestep $dt=5\times10^{-3} \omega_p^{-1} $, for plasma frequency determined from the ionization of the 2TTCP model. The particles were thermalized in an NVT Nose-Hoover thermostat for a time of $50 \omega_p^{-1}$ before a $500 \omega_p^{-1}$ NVE production phase. Thermal conductivity and viscosity were computed from the asymptotic value of time-integrated heat-flux and pressure tensor autocorrelation functions using the corresponding Green-Kubo relations.      

In Fig.~\ref{fig:Al_fixed_Ti_structure} we compare the results of DFT-MD (black pluses) to the 2TTCP model with bridge function\cite{rosenfeld1979theory} (solid blue), MD simulation (green triangles) and find close agreement amongst the three. At LTE conditions in panel $a)$, we have a strongly coupled aluminum plasma with average ionization $\langle Z \rangle=3$ defined as in\cite{Starrett2014}. Around $T_e \sim 10$ eV we start to see significant ionization until in panel $d)$ where $\langle Z \rangle=4.35$.
 The MD simulation give us confidence that the bridge function is accurate here, and we can see that our model mimics the structure of DFT-MD quite well. We additionally show the Yukawa one component plasma (YOCP) model for $T_e$ dependent $\langle Z \rangle$ from the More Thomas-Fermi ionization fit in\cite{more1985pressure} and Thomas-Fermi screening length, as well as the one component plasma (OCP) model. The OCP and YOCP curves are obtained from HNC with bridge functions \cite{PhysRevA.46.1051} and \cite{PhysRevE.61.2129}, respectively. 

\subsection{Transport Coefficients}

\begin{figure}[h]
    \centering
    \includegraphics[width=1\linewidth]{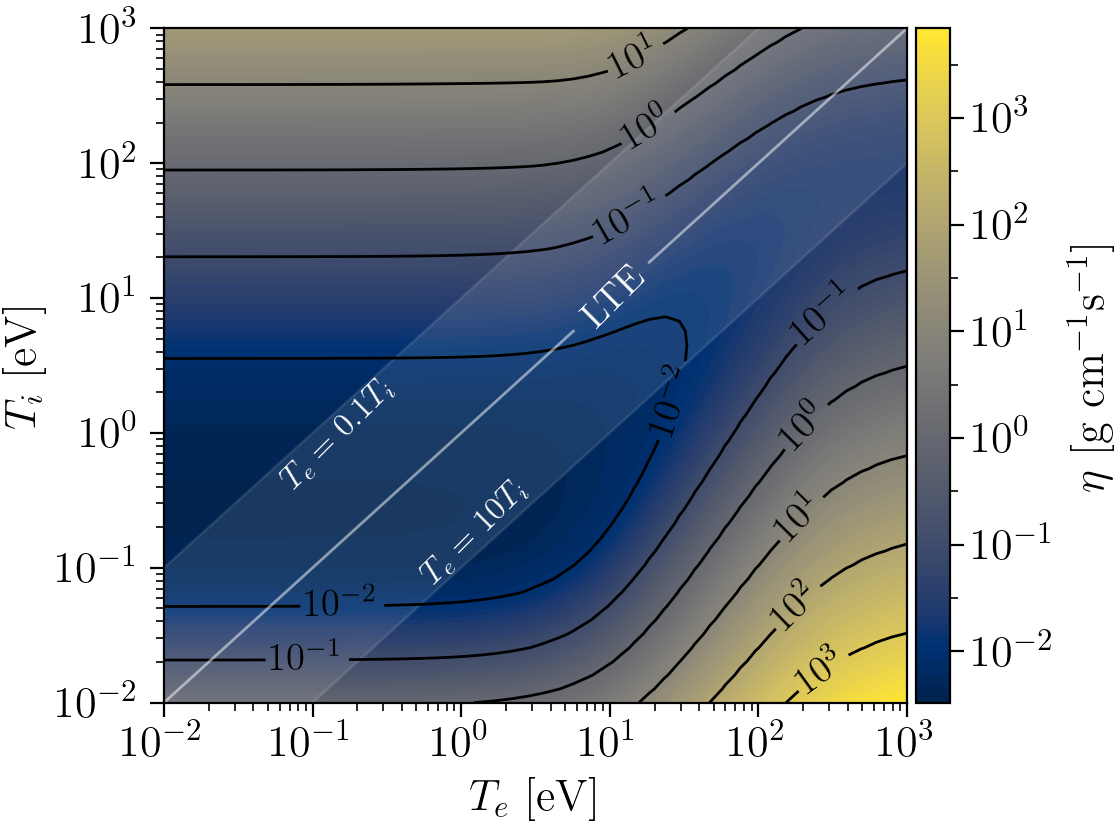}
    \caption{The viscosity of $2.7$ g/cm$^3$ aluminum from the IYVM \cite{Johnson2024}, which is refitted from \cite{MURILLO200849} to better match high temperature scaling.  } 
    \label{fig:YVM_plot}
\end{figure}

The transport properties of plasmas can depend strongly on both electron and ion temperature. In Fig.~\ref{fig:YVM_plot} we see from the improved yukawa viscosity model (IYVM) \cite{Johnson2024} that the viscosity of aluminum depends strongly on both ionic and electronic temperatures, and deviations from LTE can result in orders of magnitude of deviation from the expected answer.
The IYVM is based on fits to MD \cite{MURILLO200849, Johnson2024} that incorporate electron temperature only through a linear Thomas-Fermi screening length, and ion temperature through pair-potential MD. In the cold electron limit, the Thomas-Fermi screening length depends only on the Fermi energy and not on $T_e$, so only $T_I$ impacts viscosity until $T_e \sim {\rm few\ eV}$. As the plasma heats, variations in the pair potential from electron heating substantially affects the resulting viscosity. At high ionic temperatures the ions are kinetic and the long mean free path leads to a high viscosity, as seen in the top left of Fig.~\ref{fig:YVM_plot}. Interestingly, at high electronic temperature and low ionic temperature, the ions are highly caged from the large potential interaction, also leading to a large viscosity.

\begin{figure}[t!]
    \centering
    \includegraphics[width=1\linewidth]{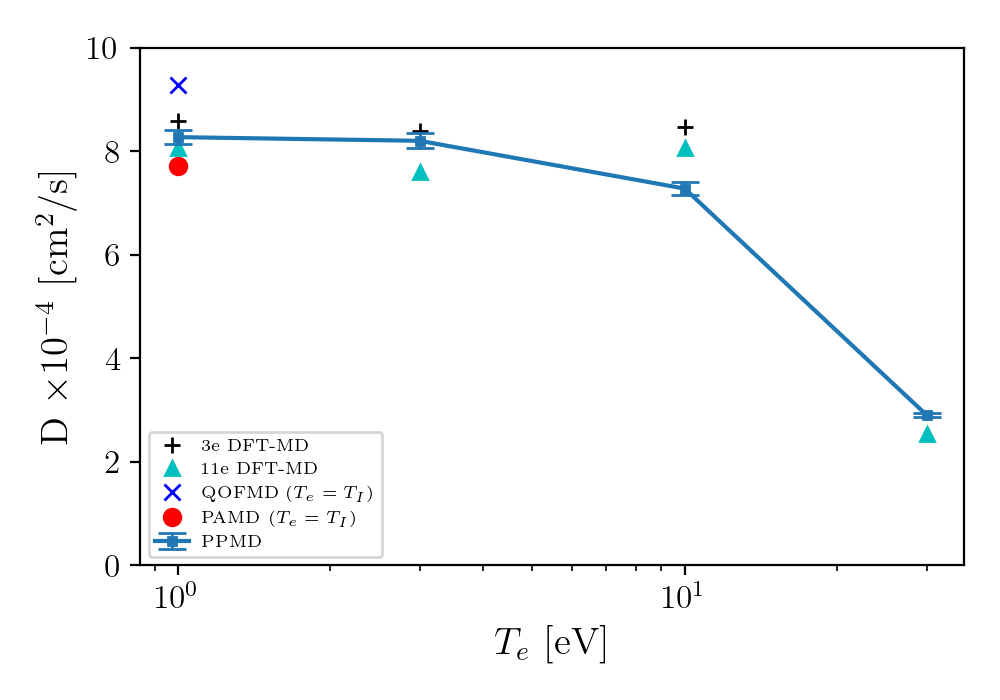}
    \caption{Aluminum self diffusion at 2.7 $g/$cm$^3$ for varying electron temperature at  $T_I=1$ eV extracted from our PPMD simulations (squares with error bars), our DFT-MD runs with three (black pluses) and eleven (cyan triangles) valence electrons. Also shown are the equilibrium models QOFMD (blue crosses) and PAMD (red circles) from \cite{PhysRevLett.116.075002}. }
    \label{fig:D_1eV}
\end{figure}

\begin{figure}[t!]
    \centering
    \includegraphics[width=1\linewidth]{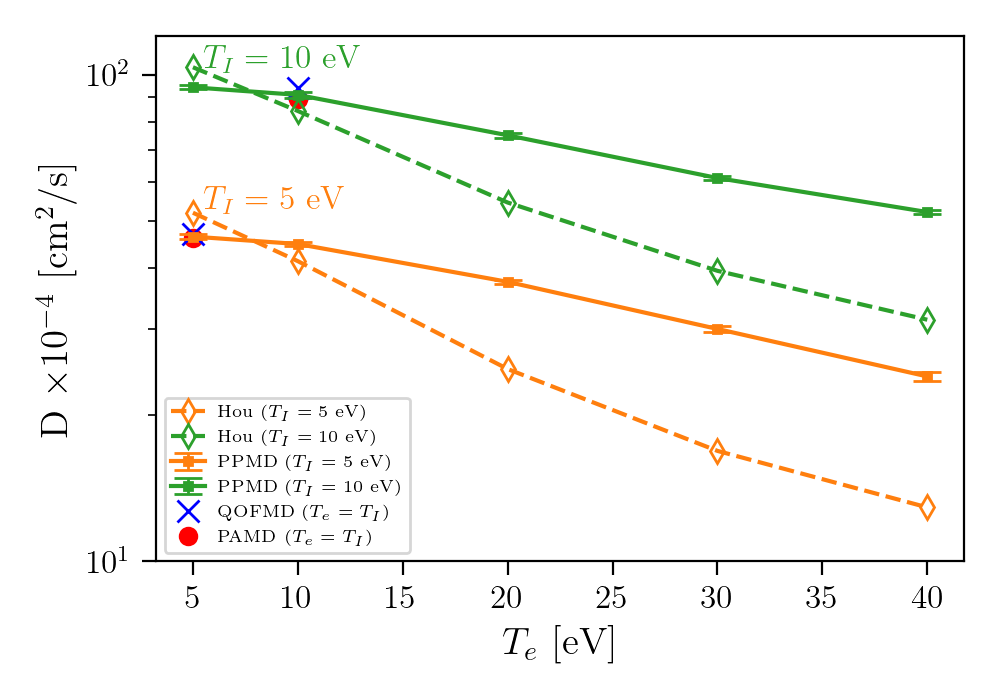}
    \caption{Aluminum self diffusion at 2.7 $g/$cm$^3$ for varying electron temperature at two different ion temperatures, $5$ eV (orange) and $10$ eV (green) extracted from our PPMD simulations (squares with error bars). Also shown is an AA model from \cite{HOU201721} (open diamonds) and the equilibrium models QOFMD (blue crosses) and PAMD (red circles) from \cite{PhysRevLett.116.075002}. }
    \label{fig:D_5-10eV}
\end{figure}

In Fig.~\ref{fig:D_1eV} we show the ion self-diffusion coefficient computed using PPMD (squares with error bars) mean square displacements\cite{HansenMcdonald} as a function of electron temperature over three different ion temperatures. Error bars here are defined as the standard deviation of the mean between the ten parallel runs of the time-integrated correlation functions through the Green-Kubo relations. Increasing electronic temperature ionizes electrons above about $10$ eV. The ion charge then increases and the cross section increases, lowering the mean free path and $D$. For the $T_I=1$ eV case, we ran two-temperature DFT-MD simulations, both for an eleven electron pseudopotential with VASP (triangles), and a three electron pseudopotential (pluses) with QuantumESPRESSO (QE) \cite{Giannozzi_2009,Giannozzi_2017}, computing the self-diffusion coefficient from mean square displacement. We find very close agreement as expected from the close agreement of the pair correlations in Fig.~\ref{fig:Al_fixed_Ti_structure}. Initial DFT-MD runs were conducted with QE with a three valence electron pseudopotential (3e), which is inadequate for Aluminum above $10$ eV, though we can see good agreement for lower T. For each of the ion temperatures, we have equilibrium results to compare to from pseudo atom molecular dynamics (PAMD) and orbital free MD (QOFMD) \cite{PhysRevLett.116.075002}\footnote{We do not plot a main result of \cite{PhysRevLett.116.075002}, the effective potential (EPT) result, simply because it overlaps the other two so closely it obscures the data}.

The only other two-temperature computation of self-diffusion to the authors knowledge is in \cite{HOU201721}, which are shown in Fig.~\ref{fig:D_5-10eV}. The AA model(diamonds) considered by Hou et al. in that paper offers an alternative two-temperature AA scheme, based on a Gordon-Kim method for generating pair potentials\cite{gordon1972theory, PhysRevE.79.016402}. This method involves a sum of overlapping atomic electron clouds which is a more approximate treatment of the free electron distribution than our Ornstein-Zernike based theory. Indeed we see better agreement with the results of \cite{PhysRevLett.116.075002}, though the difference is only around $10\%$. Additionally, our model would predict that for $T_e \lesssim 10$ eV, the ion-ion pair potential doesn't significantly change and thus neither does any ionic transport property, whereas Hou et al. data changes significantly in this range, which was also seen for the equation of state examined in \cite{PhysRevE.79.016402}. As we move to hotter electron temperatures, away from the equilibrium case, we also see increased deviation between our PPMD result and that of Hou et al. We suspect this is due to the increasing number of free electrons which are treated only approximately in that paper.    

\begin{figure}[t!]
    \centering
    \includegraphics[width=1\linewidth]{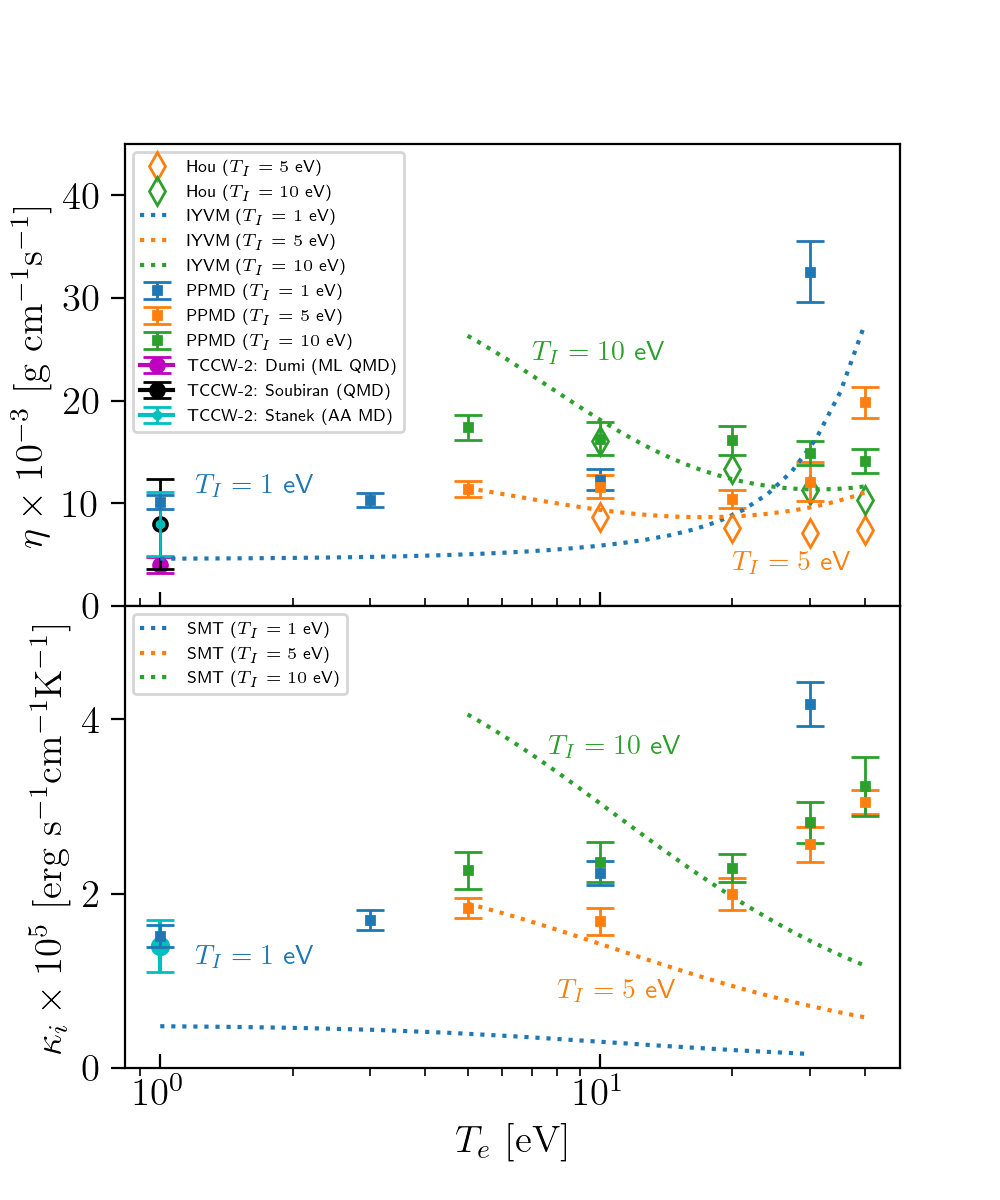}
    \caption{\text{\bf Upper)} Aluminum ion viscosity at 2.7 $g/$cm$^3$ for varying electron temperature at three different ion temperatures, $1$ eV(blue), $5$ eV orange, $10$ eV (green) extracted from our PPMD simulations (square error bars), compared with Hou et al. \cite{HOU201721} (open diamonds), equilibrium results from the TCCW2 workship\cite{Stanek2024} (purple, black, and cyan error bars), and the IYVM model (dotted lines). \text{\bf Lower)} Aluminum ion thermal conductivity for the same conditions. Here the dotted lines are based on the SMT model, rather than the IYVM.}
    \label{fig:η_κ_compare}
\end{figure}

We also compute viscosity in the same electron and ion temperature ranges, shown in the upper panel of Fig.~\ref{fig:η_κ_compare}. Here and for the below conductivity plot, the error bars correspond to the standard deviation of the mean across the three independent tensor components for each of the ten parallel runs. We again compare our PPMD result to Hou et al. finding rough agreement, but complicated temperature dependencies. The largest difference is in the highest electron temperature limit, moving far away from the dark blue viscosity minimum in Fig.~\ref{fig:YVM_plot}. The reason for the increased complexity of the viscosity compared to self-diffusion is that we care about much more than just mean free paths, and instead the potential interactions of nearby ions, a strong function of $T_e$ through $\langle Z \rangle$, is of crucial importance to the stress autocorrelation function. In the range of electron and ion temperatures shown, the total variation in viscosity appears of similar magnitude to the variation in uncertainty of the equilibrium viscosity results reported in the second transport coefficient comparison workshop (TCCW2) particles\cite{Stanek2024}. These results include an AA based PPMD (cyan) as well as two direct high fidelity DFT-MD computations. 


We finally compute ionic thermal conductivity, in the lower panel of Fig.~\ref{fig:η_κ_compare} using the heat flux autocorrelation function\cite{HansenMcdonald}. Less data is available in the literature, because the electron thermal conductivity is typically dominant. However, it is worth noting the per-species thermal conductivity scales roughly as $T^{5/2}$, meaning a non-equilibrium scenario where for example $T_I \gg T_e$, such as occurs in shocks \cite{drake2006introduction} and ICF hotstpots, could cause ion thermal conductivity to be of equal or greater importance, particularly for low Z atoms. In this hot ion limit the adiabatic approximation should still be quite accurate as long as the electron-ion temperature ratio is significantly larger than the electron-ion mass ratio.

For thermal conductivity, we compare our results with that of the SMT model (dotted lines) \cite{PhysRevE.93.043203} which is rooted in gas kinetic theory \cite{Chapman1991-as}. The binary-collision assumption can lead to qualitative disagreement in the strong coupling regime, as we see in the lower panel of Fig.~\ref{fig:η_κ_compare}. In particular, increasing electron temperatures increases the charge and thus the naive cross section---decreasing the mean free path---resulting in the decreasing thermal conductivity shown. It isn't shown, but SMT would show the same trend for viscosity in the panel above since they are exactly proportional at the level of the SMT model. This trend is wrong in both cases because SMT misses true strong coupling effects where the relevant interaction is not collisions but continuous potential interactions. Instead, our 2TTCP model coupled to PPMD is precisely the rapid computational tool needed for future exploration or fits.  


\section{Conclusions and Future Outlook}
\label{sec:Discussion}
The multi-temperature quantum Ornstein-Zernike equations we have derived, and the accompanying average atom based two-temperature two-component plasma model have been shown to accurately generalize the equilibrium answers. Since we rely on statistical mechanical techniques, albeit generalized to multiple temperatures, there are some clear limits to this calculation. One is dynamical electron effects at timescales around the electron-ion equilibration time where this model would likely be a worse approximation than at shorter and longer timescales. It is also clear that the resulting momenta distributions inherent in our model are those of a perfectly equilibrated plasma. In reality, the transient non-equilibrium system and velocity dependent cross sections would create distributions with distinct non-equilibrium distortions, whose effect on transport and EOS is not analyzed here.

This model ignores the possibility of a non-trivial cross-species temperature, such as in \cite{PhysRevE.77.026401}. For electron-ion systems, this effect is vanishingly small unless the ion temperature is orders of magnitude higher than the electron temperature, for which the Born-Oppenheimer approximation would be poor regardless. For similar mass multi-temperature ion-ion or electron-positron systems, corrections could be important. 

In this work we focused on modeling ionic structure and transport in the two-temperature limit. The 2TTCP model also offers an opportunity to model multi-temperature effects on electronic transport coefficients and equation of state properties for out of equilibrium plasmas. In this case, it is possible that fully resolving the effects of ions on primarily electron physics based properties might additionally require incorporating self-consistency such as going beyond the approximate definition of $n^{\rm scr}$. We leave this for future work.

We note that there was discussion \cite{dharma2017isochoric, DWcomment2019} on the validity of results based on the two temperature NPA. The 2TTCP model presented here relies on the same crucial equation relating the electron and ion species, Eq.~\eqref{eq:Ueff_from_nscr}, but using the model of \cite{Starrett2014}. The rigorous derivation of this equation, and our careful DFT-MD validation runs, confirms the use of this non-equilibrium model itself, independent of possible discrepancies between \cite{dharma2017isochoric,DWcomment2019}. Lastly, we note that neither of these models fully incorporates the self-consistent set of QOZ/HNC equations, the effects of which we leave to future work.

\section*{Data Availability}
    The data that support the findings of this study are available from the corresponding author upon reasonable request.

\section*{Author Contributions}
ZJ and NS conceptualized the study. ZJ developed the methodology and wrote the original draft. ZJ and NS contributed equally to the simulations. NS and MM contributed equally to manuscript review and editing.

\begin{acknowledgments}
This material is based upon work supported by the Department of Energy [National Nuclear Security Administration] University of Rochester "National Inertial Confinement Fusion Program" under Award Number(s) DE-NA0004144. Additionally, this work was supported in part through computational resources and services provided by the Institute for Cyber-Enabled Research at Michigan State University.
\end{acknowledgments}

\bibliography{bib}

\begin{thebibliography}{57}%
\makeatletter
\providecommand \@ifxundefined [1]{%
 \@ifx{#1\undefined}
}%
\providecommand \@ifnum [1]{%
 \ifnum #1\expandafter \@firstoftwo
 \else \expandafter \@secondoftwo
 \fi
}%
\providecommand \@ifx [1]{%
 \ifx #1\expandafter \@firstoftwo
 \else \expandafter \@secondoftwo
 \fi
}%
\providecommand \natexlab [1]{#1}%
\providecommand \enquote  [1]{``#1''}%
\providecommand \bibnamefont  [1]{#1}%
\providecommand \bibfnamefont [1]{#1}%
\providecommand \citenamefont [1]{#1}%
\providecommand \href@noop [0]{\@secondoftwo}%
\providecommand \href [0]{\begingroup \@sanitize@url \@href}%
\providecommand \@href[1]{\@@startlink{#1}\@@href}%
\providecommand \@@href[1]{\endgroup#1\@@endlink}%
\providecommand \@sanitize@url [0]{\catcode `\\12\catcode `\$12\catcode `\&12\catcode `\#12\catcode `\^12\catcode `\_12\catcode `\%12\relax}%
\providecommand \@@startlink[1]{}%
\providecommand \@@endlink[0]{}%
\providecommand \url  [0]{\begingroup\@sanitize@url \@url }%
\providecommand \@url [1]{\endgroup\@href {#1}{\urlprefix }}%
\providecommand \urlprefix  [0]{URL }%
\providecommand \Eprint [0]{\href }%
\providecommand \doibase [0]{https://doi.org/}%
\providecommand \selectlanguage [0]{\@gobble}%
\providecommand \bibinfo  [0]{\@secondoftwo}%
\providecommand \bibfield  [0]{\@secondoftwo}%
\providecommand \translation [1]{[#1]}%
\providecommand \BibitemOpen [0]{}%
\providecommand \bibitemStop [0]{}%
\providecommand \bibitemNoStop [0]{.\EOS\space}%
\providecommand \EOS [0]{\spacefactor3000\relax}%
\providecommand \BibitemShut  [1]{\csname bibitem#1\endcsname}%
\let\auto@bib@innerbib\@empty
\bibitem [{\citenamefont {Marinak}\ \emph {et~al.}(2001)\citenamefont {Marinak}, \citenamefont {Kerbel}, \citenamefont {Gentile}, \citenamefont {Jones}, \citenamefont {Munro}, \citenamefont {Pollaine}, \citenamefont {Dittrich},\ and\ \citenamefont {Haan}}]{Marinak2001}%
  \BibitemOpen
  \bibfield  {author} {\bibinfo {author} {\bibfnamefont {M.~M.}\ \bibnamefont {Marinak}}, \bibinfo {author} {\bibfnamefont {G.~D.}\ \bibnamefont {Kerbel}}, \bibinfo {author} {\bibfnamefont {N.~A.}\ \bibnamefont {Gentile}}, \bibinfo {author} {\bibfnamefont {O.}~\bibnamefont {Jones}}, \bibinfo {author} {\bibfnamefont {D.}~\bibnamefont {Munro}}, \bibinfo {author} {\bibfnamefont {S.}~\bibnamefont {Pollaine}}, \bibinfo {author} {\bibfnamefont {T.~R.}\ \bibnamefont {Dittrich}},\ and\ \bibinfo {author} {\bibfnamefont {S.~W.}\ \bibnamefont {Haan}},\ }\bibfield  {title} {\bibinfo {title} {Three-dimensional hydra simulations of national ignition facility targets},\ }\href {https://doi.org/10.1063/1.1356740} {\bibfield  {journal} {\bibinfo  {journal} {Physics of Plasmas}\ }\textbf {\bibinfo {volume} {8}},\ \bibinfo {pages} {2275} (\bibinfo {year} {2001})}\BibitemShut {NoStop}%
\bibitem [{\citenamefont {Fryxell}\ \emph {et~al.}(2000)\citenamefont {Fryxell}, \citenamefont {Olson}, \citenamefont {Ricker}, \citenamefont {Timmes}, \citenamefont {Zingale}, \citenamefont {Lamb}, \citenamefont {MacNeice}, \citenamefont {Rosner}, \citenamefont {Truran},\ and\ \citenamefont {Tufo}}]{Fryxell2000}%
  \BibitemOpen
  \bibfield  {author} {\bibinfo {author} {\bibfnamefont {B.}~\bibnamefont {Fryxell}}, \bibinfo {author} {\bibfnamefont {K.}~\bibnamefont {Olson}}, \bibinfo {author} {\bibfnamefont {P.}~\bibnamefont {Ricker}}, \bibinfo {author} {\bibfnamefont {F.~X.}\ \bibnamefont {Timmes}}, \bibinfo {author} {\bibfnamefont {M.}~\bibnamefont {Zingale}}, \bibinfo {author} {\bibfnamefont {D.~Q.}\ \bibnamefont {Lamb}}, \bibinfo {author} {\bibfnamefont {P.}~\bibnamefont {MacNeice}}, \bibinfo {author} {\bibfnamefont {R.}~\bibnamefont {Rosner}}, \bibinfo {author} {\bibfnamefont {J.~W.}\ \bibnamefont {Truran}},\ and\ \bibinfo {author} {\bibfnamefont {H.}~\bibnamefont {Tufo}},\ }\bibfield  {title} {\bibinfo {title} {Flash: An adaptive mesh hydrodynamics code for modeling astrophysical thermonuclear flashes},\ }\href {https://doi.org/10.1086/317361/FULLTEXT/} {\bibfield  {journal} {\bibinfo  {journal} {The Astrophysical Journal Supplement Series}\ }\textbf {\bibinfo {volume} {131}},\ \bibinfo {pages} {273} (\bibinfo {year}
  {2000})}\BibitemShut {NoStop}%
\bibitem [{\citenamefont {Orban}\ \emph {et~al.}(2013)\citenamefont {Orban}, \citenamefont {Fatenejad}, \citenamefont {Chawla}, \citenamefont {Wilks},\ and\ \citenamefont {Lamb}}]{Orban2013}%
  \BibitemOpen
  \bibfield  {author} {\bibinfo {author} {\bibfnamefont {C.}~\bibnamefont {Orban}}, \bibinfo {author} {\bibfnamefont {M.}~\bibnamefont {Fatenejad}}, \bibinfo {author} {\bibfnamefont {S.}~\bibnamefont {Chawla}}, \bibinfo {author} {\bibfnamefont {S.~C.}\ \bibnamefont {Wilks}},\ and\ \bibinfo {author} {\bibfnamefont {D.~Q.}\ \bibnamefont {Lamb}},\ }\bibfield  {title} {\bibinfo {title} {A radiation-hydrodynamics code comparison for laser-produced plasmas: Flash versus hydra and the results of validation experiments},\ }\href@noop {} {\bibfield  {journal} {\bibinfo  {journal} {arXiv preprint arXiv:1306.1584}\ } (\bibinfo {year} {2013})}\BibitemShut {NoStop}%
\bibitem [{\citenamefont {Fletcher}\ \emph {et~al.}(2022)\citenamefont {Fletcher}, \citenamefont {Vorberger}, \citenamefont {Schumaker}, \citenamefont {Ruyer}, \citenamefont {Goede}, \citenamefont {Galtier}, \citenamefont {Zastrau}, \citenamefont {Alves}, \citenamefont {Baalrud}, \citenamefont {Baggott}, \citenamefont {Barbrel}, \citenamefont {Chen}, \citenamefont {Döppner}, \citenamefont {Gauthier}, \citenamefont {Granados}, \citenamefont {Kim}, \citenamefont {Kraus}, \citenamefont {Lee}, \citenamefont {MacDonald}, \citenamefont {Mishra}, \citenamefont {Pelka}, \citenamefont {Ravasio}, \citenamefont {Roedel}, \citenamefont {Fry}, \citenamefont {Redmer}, \citenamefont {Fiuza}, \citenamefont {Gericke},\ and\ \citenamefont {Glenzer}}]{Fletcher2022}%
  \BibitemOpen
  \bibfield  {author} {\bibinfo {author} {\bibfnamefont {L.~B.}\ \bibnamefont {Fletcher}}, \bibinfo {author} {\bibfnamefont {J.}~\bibnamefont {Vorberger}}, \bibinfo {author} {\bibfnamefont {W.}~\bibnamefont {Schumaker}}, \bibinfo {author} {\bibfnamefont {C.}~\bibnamefont {Ruyer}}, \bibinfo {author} {\bibfnamefont {S.}~\bibnamefont {Goede}}, \bibinfo {author} {\bibfnamefont {E.}~\bibnamefont {Galtier}}, \bibinfo {author} {\bibfnamefont {U.}~\bibnamefont {Zastrau}}, \bibinfo {author} {\bibfnamefont {E.~P.}\ \bibnamefont {Alves}}, \bibinfo {author} {\bibfnamefont {S.~D.}\ \bibnamefont {Baalrud}}, \bibinfo {author} {\bibfnamefont {R.~A.}\ \bibnamefont {Baggott}}, \bibinfo {author} {\bibfnamefont {B.}~\bibnamefont {Barbrel}}, \bibinfo {author} {\bibfnamefont {Z.}~\bibnamefont {Chen}}, \bibinfo {author} {\bibfnamefont {T.}~\bibnamefont {Döppner}}, \bibinfo {author} {\bibfnamefont {M.}~\bibnamefont {Gauthier}}, \bibinfo {author} {\bibfnamefont {E.}~\bibnamefont {Granados}}, \bibinfo {author} {\bibfnamefont {J.~B.}\
  \bibnamefont {Kim}}, \bibinfo {author} {\bibfnamefont {D.}~\bibnamefont {Kraus}}, \bibinfo {author} {\bibfnamefont {H.~J.}\ \bibnamefont {Lee}}, \bibinfo {author} {\bibfnamefont {M.~J.}\ \bibnamefont {MacDonald}}, \bibinfo {author} {\bibfnamefont {R.}~\bibnamefont {Mishra}}, \bibinfo {author} {\bibfnamefont {A.}~\bibnamefont {Pelka}}, \bibinfo {author} {\bibfnamefont {A.}~\bibnamefont {Ravasio}}, \bibinfo {author} {\bibfnamefont {C.}~\bibnamefont {Roedel}}, \bibinfo {author} {\bibfnamefont {A.~R.}\ \bibnamefont {Fry}}, \bibinfo {author} {\bibfnamefont {R.}~\bibnamefont {Redmer}}, \bibinfo {author} {\bibfnamefont {F.}~\bibnamefont {Fiuza}}, \bibinfo {author} {\bibfnamefont {D.~O.}\ \bibnamefont {Gericke}},\ and\ \bibinfo {author} {\bibfnamefont {S.~H.}\ \bibnamefont {Glenzer}},\ }\bibfield  {title} {\bibinfo {title} {Electron-ion temperature relaxation in warm dense hydrogen observed with picosecond resolved x-ray scattering},\ }\href {https://doi.org/10.3389/FPHY.2022.838524/BIBTEX} {\bibfield  {journal}
  {\bibinfo  {journal} {Frontiers in Physics}\ }\textbf {\bibinfo {volume} {10}},\ \bibinfo {pages} {838524} (\bibinfo {year} {2022})}\BibitemShut {NoStop}%
\bibitem [{\citenamefont {Jourdain}\ \emph {et~al.}(2021)\citenamefont {Jourdain}, \citenamefont {Lecherbourg}, \citenamefont {Recoules}, \citenamefont {Renaudin},\ and\ \citenamefont {Dorchies}}]{PhysRevLett.126.065001}%
  \BibitemOpen
  \bibfield  {author} {\bibinfo {author} {\bibfnamefont {N.}~\bibnamefont {Jourdain}}, \bibinfo {author} {\bibfnamefont {L.}~\bibnamefont {Lecherbourg}}, \bibinfo {author} {\bibfnamefont {V.}~\bibnamefont {Recoules}}, \bibinfo {author} {\bibfnamefont {P.}~\bibnamefont {Renaudin}},\ and\ \bibinfo {author} {\bibfnamefont {F.}~\bibnamefont {Dorchies}},\ }\bibfield  {title} {\bibinfo {title} {Ultrafast thermal melting in nonequilibrium warm dense copper},\ }\href {https://doi.org/10.1103/PhysRevLett.126.065001} {\bibfield  {journal} {\bibinfo  {journal} {Phys. Rev. Lett.}\ }\textbf {\bibinfo {volume} {126}},\ \bibinfo {pages} {065001} (\bibinfo {year} {2021})}\BibitemShut {NoStop}%
\bibitem [{\citenamefont {Mazevet}\ \emph {et~al.}(2005)\citenamefont {Mazevet}, \citenamefont {Cl\'erouin}, \citenamefont {Recoules}, \citenamefont {Anglade},\ and\ \citenamefont {Zerah}}]{PhysRevLett.95.085002}%
  \BibitemOpen
  \bibfield  {author} {\bibinfo {author} {\bibfnamefont {S.}~\bibnamefont {Mazevet}}, \bibinfo {author} {\bibfnamefont {J.}~\bibnamefont {Cl\'erouin}}, \bibinfo {author} {\bibfnamefont {V.}~\bibnamefont {Recoules}}, \bibinfo {author} {\bibfnamefont {P.~M.}\ \bibnamefont {Anglade}},\ and\ \bibinfo {author} {\bibfnamefont {G.}~\bibnamefont {Zerah}},\ }\bibfield  {title} {\bibinfo {title} {Ab-initio simulations of the optical properties of warm dense gold},\ }\href {https://doi.org/10.1103/PhysRevLett.95.085002} {\bibfield  {journal} {\bibinfo  {journal} {Phys. Rev. Lett.}\ }\textbf {\bibinfo {volume} {95}},\ \bibinfo {pages} {085002} (\bibinfo {year} {2005})}\BibitemShut {NoStop}%
\bibitem [{\citenamefont {Lin}\ \emph {et~al.}(2008)\citenamefont {Lin}, \citenamefont {Zhigilei},\ and\ \citenamefont {Celli}}]{PhysRevB.77.075133}%
  \BibitemOpen
  \bibfield  {author} {\bibinfo {author} {\bibfnamefont {Z.}~\bibnamefont {Lin}}, \bibinfo {author} {\bibfnamefont {L.~V.}\ \bibnamefont {Zhigilei}},\ and\ \bibinfo {author} {\bibfnamefont {V.}~\bibnamefont {Celli}},\ }\bibfield  {title} {\bibinfo {title} {Electron-phonon coupling and electron heat capacity of metals under conditions of strong electron-phonon nonequilibrium},\ }\href {https://doi.org/10.1103/PhysRevB.77.075133} {\bibfield  {journal} {\bibinfo  {journal} {Phys. Rev. B}\ }\textbf {\bibinfo {volume} {77}},\ \bibinfo {pages} {075133} (\bibinfo {year} {2008})}\BibitemShut {NoStop}%
\bibitem [{\citenamefont {Wilson}\ \emph {et~al.}(2006)\citenamefont {Wilson}, \citenamefont {Sonnad}, \citenamefont {Sterne},\ and\ \citenamefont {Isaacs}}]{Wilson2006}%
  \BibitemOpen
  \bibfield  {author} {\bibinfo {author} {\bibfnamefont {B.}~\bibnamefont {Wilson}}, \bibinfo {author} {\bibfnamefont {V.}~\bibnamefont {Sonnad}}, \bibinfo {author} {\bibfnamefont {P.}~\bibnamefont {Sterne}},\ and\ \bibinfo {author} {\bibfnamefont {W.}~\bibnamefont {Isaacs}},\ }\bibfield  {title} {\bibinfo {title} {Purgatorio - a new implementation of the inferno algorithm},\ }\href {https://doi.org/10.1016/J.JQSRT.2005.05.053} {\bibfield  {journal} {\bibinfo  {journal} {Journal of Quantitative Spectroscopy and Radiative Transfer}\ }\textbf {\bibinfo {volume} {99}},\ \bibinfo {pages} {658} (\bibinfo {year} {2006})}\BibitemShut {NoStop}%
\bibitem [{\citenamefont {Starrett}\ \emph {et~al.}(2019)\citenamefont {Starrett}, \citenamefont {Gill}, \citenamefont {Sjostrom},\ and\ \citenamefont {Greeff}}]{Starrett2019}%
  \BibitemOpen
  \bibfield  {author} {\bibinfo {author} {\bibfnamefont {C.~E.}\ \bibnamefont {Starrett}}, \bibinfo {author} {\bibfnamefont {N.~M.}\ \bibnamefont {Gill}}, \bibinfo {author} {\bibfnamefont {T.}~\bibnamefont {Sjostrom}},\ and\ \bibinfo {author} {\bibfnamefont {C.~W.}\ \bibnamefont {Greeff}},\ }\bibfield  {title} {\bibinfo {title} {Wide ranging equation of state with tartarus: A hybrid green's function/orbital based average atom code},\ }\href {https://doi.org/10.1016/J.CPC.2018.10.002} {\bibfield  {journal} {\bibinfo  {journal} {Computer Physics Communications}\ }\textbf {\bibinfo {volume} {235}},\ \bibinfo {pages} {50} (\bibinfo {year} {2019})}\BibitemShut {NoStop}%
\bibitem [{\citenamefont {Starrett}\ and\ \citenamefont {Saumon}(2014)}]{Starrett2014}%
  \BibitemOpen
  \bibfield  {author} {\bibinfo {author} {\bibfnamefont {C.~E.}\ \bibnamefont {Starrett}}\ and\ \bibinfo {author} {\bibfnamefont {D.}~\bibnamefont {Saumon}},\ }\bibfield  {title} {\bibinfo {title} {A simple method for determining the ionic structure of warm dense matter},\ }\href {https://doi.org/10.1016/J.HEDP.2013.12.001} {\bibfield  {journal} {\bibinfo  {journal} {High Energy Density Physics}\ }\textbf {\bibinfo {volume} {10}},\ \bibinfo {pages} {35} (\bibinfo {year} {2014})}\BibitemShut {NoStop}%
\bibitem [{\citenamefont {Chihara}(2015)}]{Chihara2015}%
  \BibitemOpen
  \bibfield  {author} {\bibinfo {author} {\bibfnamefont {J.}~\bibnamefont {Chihara}},\ }\bibfield  {title} {\bibinfo {title} {Average atom model based on quantum hyper-netted chain method},\ }\href {https://doi.org/10.1016/j.hedp.2016.03.002} {\bibfield  {journal} {\bibinfo  {journal} {High Energy Density Physics}\ }\textbf {\bibinfo {volume} {19}},\ \bibinfo {pages} {38} (\bibinfo {year} {2015})}\BibitemShut {NoStop}%
\bibitem [{\citenamefont {Massacrier}\ \emph {et~al.}(2021)\citenamefont {Massacrier}, \citenamefont {B\"ohme}, \citenamefont {Vorberger}, \citenamefont {Soubiran},\ and\ \citenamefont {Militzer}}]{PhysRevResearch.3.023026}%
  \BibitemOpen
  \bibfield  {author} {\bibinfo {author} {\bibfnamefont {G.}~\bibnamefont {Massacrier}}, \bibinfo {author} {\bibfnamefont {M.}~\bibnamefont {B\"ohme}}, \bibinfo {author} {\bibfnamefont {J.}~\bibnamefont {Vorberger}}, \bibinfo {author} {\bibfnamefont {F.}~\bibnamefont {Soubiran}},\ and\ \bibinfo {author} {\bibfnamefont {B.}~\bibnamefont {Militzer}},\ }\bibfield  {title} {\bibinfo {title} {Reconciling ionization energies and band gaps of warm dense matter derived with ab initio simulations and average atom models},\ }\href {https://doi.org/10.1103/PhysRevResearch.3.023026} {\bibfield  {journal} {\bibinfo  {journal} {Phys. Rev. Res.}\ }\textbf {\bibinfo {volume} {3}},\ \bibinfo {pages} {023026} (\bibinfo {year} {2021})}\BibitemShut {NoStop}%
\bibitem [{\citenamefont {Hansen}(2023)}]{Hansen2023}%
  \BibitemOpen
  \bibfield  {author} {\bibinfo {author} {\bibfnamefont {S.~B.}\ \bibnamefont {Hansen}},\ }\bibfield  {title} {\bibinfo {title} {Self-consistent and detailed opacities from a non-equilibrium average-atom model},\ }\bibfield  {journal} {\bibinfo  {journal} {Philosophical Transactions of the Royal Society A: Mathematical, Physical and Engineering Sciences}\ }\textbf {\bibinfo {volume} {381}},\ \href {https://doi.org/10.1098/rsta.2022.0212} {10.1098/rsta.2022.0212} (\bibinfo {year} {2023})\BibitemShut {NoStop}%
\bibitem [{\citenamefont {Callow}\ \emph {et~al.}(2022)\citenamefont {Callow}, \citenamefont {Kotik}, \citenamefont {Kraisler},\ and\ \citenamefont {Cangi}}]{callow2022atomec}%
  \BibitemOpen
  \bibfield  {author} {\bibinfo {author} {\bibfnamefont {T.~J.}\ \bibnamefont {Callow}}, \bibinfo {author} {\bibfnamefont {D.}~\bibnamefont {Kotik}}, \bibinfo {author} {\bibfnamefont {E.}~\bibnamefont {Kraisler}},\ and\ \bibinfo {author} {\bibfnamefont {A.}~\bibnamefont {Cangi}},\ }\bibfield  {title} {\bibinfo {title} {atomec: An open-source average-atom python code},\ }\href@noop {} {\bibfield  {journal} {\bibinfo  {journal} {arXiv preprint arXiv:2206.01074}\ } (\bibinfo {year} {2022})}\BibitemShut {NoStop}%
\bibitem [{\citenamefont {Piron}\ and\ \citenamefont {Blenski}(2011)}]{PhysRevE.83.026403}%
  \BibitemOpen
  \bibfield  {author} {\bibinfo {author} {\bibfnamefont {R.}~\bibnamefont {Piron}}\ and\ \bibinfo {author} {\bibfnamefont {T.}~\bibnamefont {Blenski}},\ }\bibfield  {title} {\bibinfo {title} {Variational-average-atom-in-quantum-plasmas (vaaqp) code and virial theorem: Equation-of-state and shock-hugoniot calculations for warm dense al, fe, cu, and pb},\ }\href {https://doi.org/10.1103/PhysRevE.83.026403} {\bibfield  {journal} {\bibinfo  {journal} {Phys. Rev. E}\ }\textbf {\bibinfo {volume} {83}},\ \bibinfo {pages} {026403} (\bibinfo {year} {2011})}\BibitemShut {NoStop}%
\bibitem [{\citenamefont {Grabowski}\ \emph {et~al.}(2020)\citenamefont {Grabowski}, \citenamefont {Hansen}, \citenamefont {Murillo}, \citenamefont {Stanton}, \citenamefont {Graziani}, \citenamefont {Zylstra}, \citenamefont {Baalrud}, \citenamefont {Arnault}, \citenamefont {Baczewski}, \citenamefont {Benedict}, \citenamefont {Blancard}, \citenamefont {Čertík}, \citenamefont {Clérouin}, \citenamefont {Collins}, \citenamefont {Copeland}, \citenamefont {Correa}, \citenamefont {Dai}, \citenamefont {Daligault}, \citenamefont {Desjarlais}, \citenamefont {Dharma-wardana}, \citenamefont {Faussurier}, \citenamefont {Haack}, \citenamefont {Haxhimali}, \citenamefont {Hayes-Sterbenz}, \citenamefont {Hou}, \citenamefont {Hu}, \citenamefont {Jensen}, \citenamefont {Jungman}, \citenamefont {Kagan}, \citenamefont {Kang}, \citenamefont {Kress}, \citenamefont {Ma}, \citenamefont {Marciante}, \citenamefont {Meyer}, \citenamefont {Rudd}, \citenamefont {Saumon}, \citenamefont {Shulenburger}, \citenamefont {Singleton},
  \citenamefont {Sjostrom}, \citenamefont {Stanek}, \citenamefont {Starrett}, \citenamefont {Ticknor}, \citenamefont {Valaitis}, \citenamefont {Venzke},\ and\ \citenamefont {White}}]{Grabowski2020}%
  \BibitemOpen
  \bibfield  {author} {\bibinfo {author} {\bibfnamefont {P.~E.}\ \bibnamefont {Grabowski}}, \bibinfo {author} {\bibfnamefont {S.~B.}\ \bibnamefont {Hansen}}, \bibinfo {author} {\bibfnamefont {M.~S.}\ \bibnamefont {Murillo}}, \bibinfo {author} {\bibfnamefont {L.~G.}\ \bibnamefont {Stanton}}, \bibinfo {author} {\bibfnamefont {F.~R.}\ \bibnamefont {Graziani}}, \bibinfo {author} {\bibfnamefont {A.~B.}\ \bibnamefont {Zylstra}}, \bibinfo {author} {\bibfnamefont {S.~D.}\ \bibnamefont {Baalrud}}, \bibinfo {author} {\bibfnamefont {P.}~\bibnamefont {Arnault}}, \bibinfo {author} {\bibfnamefont {A.~D.}\ \bibnamefont {Baczewski}}, \bibinfo {author} {\bibfnamefont {L.~X.}\ \bibnamefont {Benedict}}, \bibinfo {author} {\bibfnamefont {C.}~\bibnamefont {Blancard}}, \bibinfo {author} {\bibfnamefont {O.}~\bibnamefont {Čertík}}, \bibinfo {author} {\bibfnamefont {J.}~\bibnamefont {Clérouin}}, \bibinfo {author} {\bibfnamefont {L.~A.}\ \bibnamefont {Collins}}, \bibinfo {author} {\bibfnamefont {S.}~\bibnamefont {Copeland}},
  \bibinfo {author} {\bibfnamefont {A.~A.}\ \bibnamefont {Correa}}, \bibinfo {author} {\bibfnamefont {J.}~\bibnamefont {Dai}}, \bibinfo {author} {\bibfnamefont {J.}~\bibnamefont {Daligault}}, \bibinfo {author} {\bibfnamefont {M.~P.}\ \bibnamefont {Desjarlais}}, \bibinfo {author} {\bibfnamefont {M.~W.}\ \bibnamefont {Dharma-wardana}}, \bibinfo {author} {\bibfnamefont {G.}~\bibnamefont {Faussurier}}, \bibinfo {author} {\bibfnamefont {J.}~\bibnamefont {Haack}}, \bibinfo {author} {\bibfnamefont {T.}~\bibnamefont {Haxhimali}}, \bibinfo {author} {\bibfnamefont {A.}~\bibnamefont {Hayes-Sterbenz}}, \bibinfo {author} {\bibfnamefont {Y.}~\bibnamefont {Hou}}, \bibinfo {author} {\bibfnamefont {S.~X.}\ \bibnamefont {Hu}}, \bibinfo {author} {\bibfnamefont {D.}~\bibnamefont {Jensen}}, \bibinfo {author} {\bibfnamefont {G.}~\bibnamefont {Jungman}}, \bibinfo {author} {\bibfnamefont {G.}~\bibnamefont {Kagan}}, \bibinfo {author} {\bibfnamefont {D.}~\bibnamefont {Kang}}, \bibinfo {author} {\bibfnamefont {J.~D.}\ \bibnamefont
  {Kress}}, \bibinfo {author} {\bibfnamefont {Q.}~\bibnamefont {Ma}}, \bibinfo {author} {\bibfnamefont {M.}~\bibnamefont {Marciante}}, \bibinfo {author} {\bibfnamefont {E.}~\bibnamefont {Meyer}}, \bibinfo {author} {\bibfnamefont {R.~E.}\ \bibnamefont {Rudd}}, \bibinfo {author} {\bibfnamefont {D.}~\bibnamefont {Saumon}}, \bibinfo {author} {\bibfnamefont {L.}~\bibnamefont {Shulenburger}}, \bibinfo {author} {\bibfnamefont {R.~L.}\ \bibnamefont {Singleton}}, \bibinfo {author} {\bibfnamefont {T.}~\bibnamefont {Sjostrom}}, \bibinfo {author} {\bibfnamefont {L.~J.}\ \bibnamefont {Stanek}}, \bibinfo {author} {\bibfnamefont {C.~E.}\ \bibnamefont {Starrett}}, \bibinfo {author} {\bibfnamefont {C.}~\bibnamefont {Ticknor}}, \bibinfo {author} {\bibfnamefont {S.}~\bibnamefont {Valaitis}}, \bibinfo {author} {\bibfnamefont {J.}~\bibnamefont {Venzke}},\ and\ \bibinfo {author} {\bibfnamefont {A.}~\bibnamefont {White}},\ }\bibfield  {title} {\bibinfo {title} {Review of the first charged-particle transport coefficient comparison
  workshop},\ }\href {https://doi.org/10.1016/J.HEDP.2020.100905} {\bibfield  {journal} {\bibinfo  {journal} {High Energy Density Physics}\ }\textbf {\bibinfo {volume} {37}},\ \bibinfo {pages} {100905} (\bibinfo {year} {2020})}\BibitemShut {NoStop}%
\bibitem [{\citenamefont {Stanek}\ \emph {et~al.}(2024)\citenamefont {Stanek}, \citenamefont {Kononov}, \citenamefont {Hansen}, \citenamefont {Haines}, \citenamefont {Hu}, \citenamefont {Knapp}, \citenamefont {Murillo}, \citenamefont {Stanton}, \citenamefont {Whitley}, \citenamefont {Baalrud}, \citenamefont {Babati}, \citenamefont {Baczewski}, \citenamefont {Bethkenhagen}, \citenamefont {Blanchet}, \citenamefont {Clay}, \citenamefont {Cochrane}, \citenamefont {Collins}, \citenamefont {Dumi}, \citenamefont {Faussurier}, \citenamefont {French}, \citenamefont {Johnson}, \citenamefont {Karasiev}, \citenamefont {Kumar}, \citenamefont {Lentz}, \citenamefont {Melton}, \citenamefont {Nichols}, \citenamefont {Petrov}, \citenamefont {Recoules}, \citenamefont {Redmer}, \citenamefont {Röpke}, \citenamefont {Schörner}, \citenamefont {Shaffer}, \citenamefont {Sharma}, \citenamefont {Silvestri}, \citenamefont {Soubiran}, \citenamefont {Suryanarayana}, \citenamefont {Tacu}, \citenamefont {Townsend},\ and\ \citenamefont
  {White}}]{Stanek2024}%
  \BibitemOpen
  \bibfield  {author} {\bibinfo {author} {\bibfnamefont {L.~J.}\ \bibnamefont {Stanek}}, \bibinfo {author} {\bibfnamefont {A.}~\bibnamefont {Kononov}}, \bibinfo {author} {\bibfnamefont {S.~B.}\ \bibnamefont {Hansen}}, \bibinfo {author} {\bibfnamefont {B.~M.}\ \bibnamefont {Haines}}, \bibinfo {author} {\bibfnamefont {S.~X.}\ \bibnamefont {Hu}}, \bibinfo {author} {\bibfnamefont {P.~F.}\ \bibnamefont {Knapp}}, \bibinfo {author} {\bibfnamefont {M.~S.}\ \bibnamefont {Murillo}}, \bibinfo {author} {\bibfnamefont {L.~G.}\ \bibnamefont {Stanton}}, \bibinfo {author} {\bibfnamefont {H.~D.}\ \bibnamefont {Whitley}}, \bibinfo {author} {\bibfnamefont {S.~D.}\ \bibnamefont {Baalrud}}, \bibinfo {author} {\bibfnamefont {L.~J.}\ \bibnamefont {Babati}}, \bibinfo {author} {\bibfnamefont {A.~D.}\ \bibnamefont {Baczewski}}, \bibinfo {author} {\bibfnamefont {M.}~\bibnamefont {Bethkenhagen}}, \bibinfo {author} {\bibfnamefont {A.}~\bibnamefont {Blanchet}}, \bibinfo {author} {\bibfnamefont {R.~C.}\ \bibnamefont {Clay}}, \bibinfo
  {author} {\bibfnamefont {K.~R.}\ \bibnamefont {Cochrane}}, \bibinfo {author} {\bibfnamefont {L.~A.}\ \bibnamefont {Collins}}, \bibinfo {author} {\bibfnamefont {A.}~\bibnamefont {Dumi}}, \bibinfo {author} {\bibfnamefont {G.}~\bibnamefont {Faussurier}}, \bibinfo {author} {\bibfnamefont {M.}~\bibnamefont {French}}, \bibinfo {author} {\bibfnamefont {Z.~A.}\ \bibnamefont {Johnson}}, \bibinfo {author} {\bibfnamefont {V.~V.}\ \bibnamefont {Karasiev}}, \bibinfo {author} {\bibfnamefont {S.}~\bibnamefont {Kumar}}, \bibinfo {author} {\bibfnamefont {M.~K.}\ \bibnamefont {Lentz}}, \bibinfo {author} {\bibfnamefont {C.~A.}\ \bibnamefont {Melton}}, \bibinfo {author} {\bibfnamefont {K.~A.}\ \bibnamefont {Nichols}}, \bibinfo {author} {\bibfnamefont {G.~M.}\ \bibnamefont {Petrov}}, \bibinfo {author} {\bibfnamefont {V.}~\bibnamefont {Recoules}}, \bibinfo {author} {\bibfnamefont {R.}~\bibnamefont {Redmer}}, \bibinfo {author} {\bibfnamefont {G.}~\bibnamefont {Röpke}}, \bibinfo {author} {\bibfnamefont {M.}~\bibnamefont
  {Schörner}}, \bibinfo {author} {\bibfnamefont {N.~R.}\ \bibnamefont {Shaffer}}, \bibinfo {author} {\bibfnamefont {V.}~\bibnamefont {Sharma}}, \bibinfo {author} {\bibfnamefont {L.~G.}\ \bibnamefont {Silvestri}}, \bibinfo {author} {\bibfnamefont {F.}~\bibnamefont {Soubiran}}, \bibinfo {author} {\bibfnamefont {P.}~\bibnamefont {Suryanarayana}}, \bibinfo {author} {\bibfnamefont {M.}~\bibnamefont {Tacu}}, \bibinfo {author} {\bibfnamefont {J.~P.}\ \bibnamefont {Townsend}},\ and\ \bibinfo {author} {\bibfnamefont {A.~J.}\ \bibnamefont {White}},\ }\bibfield  {title} {\bibinfo {title} {Review of the second charged-particle transport coefficient code comparison workshop},\ }\href {https://doi.org/10.1063/5.0198155/3287990} {\bibfield  {journal} {\bibinfo  {journal} {Physics of Plasmas}\ }\textbf {\bibinfo {volume} {31}},\ \bibinfo {pages} {52104} (\bibinfo {year} {2024})}\BibitemShut {NoStop}%
\bibitem [{\citenamefont {Dharma-wardana}(2012)}]{DW2T-2012}%
  \BibitemOpen
  \bibfield  {author} {\bibinfo {author} {\bibfnamefont {M.~W.~C.}\ \bibnamefont {Dharma-wardana}},\ }\bibfield  {title} {\bibinfo {title} {Electron-ion and ion-ion potentials for modeling warm dense matter: Applications to laser-heated or shock-compressed al and si},\ }\href {https://doi.org/10.1103/PhysRevE.86.036407} {\bibfield  {journal} {\bibinfo  {journal} {Phys. Rev. E}\ }\textbf {\bibinfo {volume} {86}},\ \bibinfo {pages} {036407} (\bibinfo {year} {2012})}\BibitemShut {NoStop}%
\bibitem [{\citenamefont {Dharma-wardana}\ and\ \citenamefont {Perrot}(1998)}]{PhysRevE.58.3705}%
  \BibitemOpen
  \bibfield  {author} {\bibinfo {author} {\bibfnamefont {M.~W.~C.}\ \bibnamefont {Dharma-wardana}}\ and\ \bibinfo {author} {\bibfnamefont {F.~m.~c.}\ \bibnamefont {Perrot}},\ }\bibfield  {title} {\bibinfo {title} {Energy relaxation and the quasiequation of state of a dense two-temperature nonequilibrium plasma},\ }\href {https://doi.org/10.1103/PhysRevE.58.3705} {\bibfield  {journal} {\bibinfo  {journal} {Phys. Rev. E}\ }\textbf {\bibinfo {volume} {58}},\ \bibinfo {pages} {3705} (\bibinfo {year} {1998})}\BibitemShut {NoStop}%
\bibitem [{\citenamefont {Gordon}\ and\ \citenamefont {Kim}(1972)}]{gordon1972theory}%
  \BibitemOpen
  \bibfield  {author} {\bibinfo {author} {\bibfnamefont {R.~G.}\ \bibnamefont {Gordon}}\ and\ \bibinfo {author} {\bibfnamefont {Y.~S.}\ \bibnamefont {Kim}},\ }\bibfield  {title} {\bibinfo {title} {Theory for the forces between closed-shell atoms and molecules},\ }\href@noop {} {\bibfield  {journal} {\bibinfo  {journal} {The Journal of Chemical Physics}\ }\textbf {\bibinfo {volume} {56}},\ \bibinfo {pages} {3122} (\bibinfo {year} {1972})}\BibitemShut {NoStop}%
\bibitem [{\citenamefont {Hou}\ \emph {et~al.}(2017)\citenamefont {Hou}, \citenamefont {Fu}, \citenamefont {Bredow}, \citenamefont {Kang}, \citenamefont {Redmer},\ and\ \citenamefont {Yuan}}]{HOU201721}%
  \BibitemOpen
  \bibfield  {author} {\bibinfo {author} {\bibfnamefont {Y.}~\bibnamefont {Hou}}, \bibinfo {author} {\bibfnamefont {Y.}~\bibnamefont {Fu}}, \bibinfo {author} {\bibfnamefont {R.}~\bibnamefont {Bredow}}, \bibinfo {author} {\bibfnamefont {D.}~\bibnamefont {Kang}}, \bibinfo {author} {\bibfnamefont {R.}~\bibnamefont {Redmer}},\ and\ \bibinfo {author} {\bibfnamefont {J.}~\bibnamefont {Yuan}},\ }\bibfield  {title} {\bibinfo {title} {Average-atom model for two-temperature states and ionic transport properties of aluminum in the warm dense matter regime},\ }\href {https://doi.org/https://doi.org/10.1016/j.hedp.2017.01.003} {\bibfield  {journal} {\bibinfo  {journal} {High Energy Density Physics}\ }\textbf {\bibinfo {volume} {22}},\ \bibinfo {pages} {21} (\bibinfo {year} {2017})}\BibitemShut {NoStop}%
\bibitem [{\citenamefont {Boercker}\ and\ \citenamefont {More}(1986)}]{BoerckerMore1986}%
  \BibitemOpen
  \bibfield  {author} {\bibinfo {author} {\bibfnamefont {D.~B.}\ \bibnamefont {Boercker}}\ and\ \bibinfo {author} {\bibfnamefont {R.~M.}\ \bibnamefont {More}},\ }\bibfield  {title} {\bibinfo {title} {Statistical mechanics of a two-temperature, classical plasma},\ }\href {https://doi.org/10.1103/PhysRevA.33.1859} {\bibfield  {journal} {\bibinfo  {journal} {Phys. Rev. A}\ }\textbf {\bibinfo {volume} {33}},\ \bibinfo {pages} {1859} (\bibinfo {year} {1986})}\BibitemShut {NoStop}%
\bibitem [{\citenamefont {Stanek}\ \emph {et~al.}(2021)\citenamefont {Stanek}, \citenamefont {Clay}, \citenamefont {Dharma-Wardana}, \citenamefont {Wood}, \citenamefont {Beckwith},\ and\ \citenamefont {Murillo}}]{stanek2021efficacy}%
  \BibitemOpen
  \bibfield  {author} {\bibinfo {author} {\bibfnamefont {L.~J.}\ \bibnamefont {Stanek}}, \bibinfo {author} {\bibfnamefont {R.~C.}\ \bibnamefont {Clay}}, \bibinfo {author} {\bibfnamefont {M.}~\bibnamefont {Dharma-Wardana}}, \bibinfo {author} {\bibfnamefont {M.~A.}\ \bibnamefont {Wood}}, \bibinfo {author} {\bibfnamefont {K.~R.}\ \bibnamefont {Beckwith}},\ and\ \bibinfo {author} {\bibfnamefont {M.~S.}\ \bibnamefont {Murillo}},\ }\bibfield  {title} {\bibinfo {title} {Efficacy of the radial pair potential approximation for molecular dynamics simulations of dense plasmas},\ }\href@noop {} {\bibfield  {journal} {\bibinfo  {journal} {Physics of Plasmas}\ }\textbf {\bibinfo {volume} {28}} (\bibinfo {year} {2021})}\BibitemShut {NoStop}%
\bibitem [{\citenamefont {Harbour}\ \emph {et~al.}(2017)\citenamefont {Harbour}, \citenamefont {Dharma-Wardana}, \citenamefont {Klug},\ and\ \citenamefont {Lewis}}]{harbour2017equation}%
  \BibitemOpen
  \bibfield  {author} {\bibinfo {author} {\bibfnamefont {L.}~\bibnamefont {Harbour}}, \bibinfo {author} {\bibfnamefont {M.}~\bibnamefont {Dharma-Wardana}}, \bibinfo {author} {\bibfnamefont {D.~D.}\ \bibnamefont {Klug}},\ and\ \bibinfo {author} {\bibfnamefont {L.~J.}\ \bibnamefont {Lewis}},\ }\bibfield  {title} {\bibinfo {title} {Equation of state, phonons, and lattice stability of ultrafast warm dense matter},\ }\href@noop {} {\bibfield  {journal} {\bibinfo  {journal} {Physical Review E}\ }\textbf {\bibinfo {volume} {95}},\ \bibinfo {pages} {043201} (\bibinfo {year} {2017})}\BibitemShut {NoStop}%
\bibitem [{\citenamefont {Harbour}\ \emph {et~al.}(2018)\citenamefont {Harbour}, \citenamefont {F{\"o}rster}, \citenamefont {Dharma-Wardana},\ and\ \citenamefont {Lewis}}]{harbour2018ion}%
  \BibitemOpen
  \bibfield  {author} {\bibinfo {author} {\bibfnamefont {L.}~\bibnamefont {Harbour}}, \bibinfo {author} {\bibfnamefont {G.}~\bibnamefont {F{\"o}rster}}, \bibinfo {author} {\bibfnamefont {M.}~\bibnamefont {Dharma-Wardana}},\ and\ \bibinfo {author} {\bibfnamefont {L.~J.}\ \bibnamefont {Lewis}},\ }\bibfield  {title} {\bibinfo {title} {Ion-ion dynamic structure factor, acoustic modes, and equation of state of two-temperature warm dense aluminum},\ }\href@noop {} {\bibfield  {journal} {\bibinfo  {journal} {Physical review E}\ }\textbf {\bibinfo {volume} {97}},\ \bibinfo {pages} {043210} (\bibinfo {year} {2018})}\BibitemShut {NoStop}%
\bibitem [{\citenamefont {Dharma-Wardana}\ \emph {et~al.}(2017)\citenamefont {Dharma-Wardana}, \citenamefont {Klug}, \citenamefont {Harbour},\ and\ \citenamefont {Lewis}}]{dharma2017isochoric}%
  \BibitemOpen
  \bibfield  {author} {\bibinfo {author} {\bibfnamefont {M.}~\bibnamefont {Dharma-Wardana}}, \bibinfo {author} {\bibfnamefont {D.}~\bibnamefont {Klug}}, \bibinfo {author} {\bibfnamefont {L.}~\bibnamefont {Harbour}},\ and\ \bibinfo {author} {\bibfnamefont {L.~J.}\ \bibnamefont {Lewis}},\ }\bibfield  {title} {\bibinfo {title} {Isochoric, isobaric, and ultrafast conductivities of aluminum, lithium, and carbon in the warm dense matter regime},\ }\href@noop {} {\bibfield  {journal} {\bibinfo  {journal} {Physical Review E}\ }\textbf {\bibinfo {volume} {96}},\ \bibinfo {pages} {053206} (\bibinfo {year} {2017})}\BibitemShut {NoStop}%
\bibitem [{\citenamefont {Stanton}\ and\ \citenamefont {Murillo}(2016)}]{PhysRevE.93.043203}%
  \BibitemOpen
  \bibfield  {author} {\bibinfo {author} {\bibfnamefont {L.~G.}\ \bibnamefont {Stanton}}\ and\ \bibinfo {author} {\bibfnamefont {M.~S.}\ \bibnamefont {Murillo}},\ }\bibfield  {title} {\bibinfo {title} {Ionic transport in high-energy-density matter},\ }\href {https://doi.org/10.1103/PhysRevE.93.043203} {\bibfield  {journal} {\bibinfo  {journal} {Phys. Rev. E}\ }\textbf {\bibinfo {volume} {93}},\ \bibinfo {pages} {043203} (\bibinfo {year} {2016})}\BibitemShut {NoStop}%
\bibitem [{\citenamefont {Murillo}(2008)}]{MURILLO200849}%
  \BibitemOpen
  \bibfield  {author} {\bibinfo {author} {\bibfnamefont {M.~S.}\ \bibnamefont {Murillo}},\ }\bibfield  {title} {\bibinfo {title} {Viscosity estimates of liquid metals and warm dense matter using the yukawa reference system},\ }\href {https://doi.org/https://doi.org/10.1016/j.hedp.2007.11.001} {\bibfield  {journal} {\bibinfo  {journal} {High Energy Density Physics}\ }\textbf {\bibinfo {volume} {4}},\ \bibinfo {pages} {49} (\bibinfo {year} {2008})}\BibitemShut {NoStop}%
\bibitem [{\citenamefont {Johnson}\ \emph {et~al.}(2024)\citenamefont {Johnson}, \citenamefont {Silvestri}, \citenamefont {Petrov}, \citenamefont {Stanton},\ and\ \citenamefont {Murillo}}]{Johnson2024}%
  \BibitemOpen
  \bibfield  {author} {\bibinfo {author} {\bibfnamefont {Z.~A.}\ \bibnamefont {Johnson}}, \bibinfo {author} {\bibfnamefont {L.~G.}\ \bibnamefont {Silvestri}}, \bibinfo {author} {\bibfnamefont {G.~M.}\ \bibnamefont {Petrov}}, \bibinfo {author} {\bibfnamefont {L.~G.}\ \bibnamefont {Stanton}},\ and\ \bibinfo {author} {\bibfnamefont {M.~S.}\ \bibnamefont {Murillo}},\ }\bibfield  {title} {\bibinfo {title} {{Comparison of transport models in dense plasmas}},\ }\href {https://doi.org/10.1063/5.0204226} {\bibfield  {journal} {\bibinfo  {journal} {Physics of Plasmas}\ }\textbf {\bibinfo {volume} {31}},\ \bibinfo {pages} {082701} (\bibinfo {year} {2024})},\ \Eprint {https://arxiv.org/abs/https://pubs.aip.org/aip/pop/article-pdf/doi/10.1063/5.0204226/20089386/082701\_1\_5.0204226.pdf} {https://pubs.aip.org/aip/pop/article-pdf/doi/10.1063/5.0204226/20089386/082701\_1\_5.0204226.pdf} \BibitemShut {NoStop}%
\bibitem [{\citenamefont {Giannozzi}\ \emph {et~al.}(2009)\citenamefont {Giannozzi}, \citenamefont {Baroni}, \citenamefont {Bonini}, \citenamefont {Calandra}, \citenamefont {Car}, \citenamefont {Cavazzoni}, \citenamefont {Ceresoli}, \citenamefont {Chiarotti}, \citenamefont {Cococcioni}, \citenamefont {Dabo}, \citenamefont {Corso}, \citenamefont {de~Gironcoli}, \citenamefont {Fabris}, \citenamefont {Fratesi}, \citenamefont {Gebauer}, \citenamefont {Gerstmann}, \citenamefont {Gougoussis}, \citenamefont {Kokalj}, \citenamefont {Lazzeri}, \citenamefont {Martin-Samos}, \citenamefont {Marzari}, \citenamefont {Mauri}, \citenamefont {Mazzarello}, \citenamefont {Paolini}, \citenamefont {Pasquarello}, \citenamefont {Paulatto}, \citenamefont {Sbraccia}, \citenamefont {Scandolo}, \citenamefont {Sclauzero}, \citenamefont {Seitsonen}, \citenamefont {Smogunov}, \citenamefont {Umari},\ and\ \citenamefont {Wentzcovitch}}]{Giannozzi_2009}%
  \BibitemOpen
  \bibfield  {author} {\bibinfo {author} {\bibfnamefont {P.}~\bibnamefont {Giannozzi}}, \bibinfo {author} {\bibfnamefont {S.}~\bibnamefont {Baroni}}, \bibinfo {author} {\bibfnamefont {N.}~\bibnamefont {Bonini}}, \bibinfo {author} {\bibfnamefont {M.}~\bibnamefont {Calandra}}, \bibinfo {author} {\bibfnamefont {R.}~\bibnamefont {Car}}, \bibinfo {author} {\bibfnamefont {C.}~\bibnamefont {Cavazzoni}}, \bibinfo {author} {\bibfnamefont {D.}~\bibnamefont {Ceresoli}}, \bibinfo {author} {\bibfnamefont {G.~L.}\ \bibnamefont {Chiarotti}}, \bibinfo {author} {\bibfnamefont {M.}~\bibnamefont {Cococcioni}}, \bibinfo {author} {\bibfnamefont {I.}~\bibnamefont {Dabo}}, \bibinfo {author} {\bibfnamefont {A.~D.}\ \bibnamefont {Corso}}, \bibinfo {author} {\bibfnamefont {S.}~\bibnamefont {de~Gironcoli}}, \bibinfo {author} {\bibfnamefont {S.}~\bibnamefont {Fabris}}, \bibinfo {author} {\bibfnamefont {G.}~\bibnamefont {Fratesi}}, \bibinfo {author} {\bibfnamefont {R.}~\bibnamefont {Gebauer}}, \bibinfo {author} {\bibfnamefont
  {U.}~\bibnamefont {Gerstmann}}, \bibinfo {author} {\bibfnamefont {C.}~\bibnamefont {Gougoussis}}, \bibinfo {author} {\bibfnamefont {A.}~\bibnamefont {Kokalj}}, \bibinfo {author} {\bibfnamefont {M.}~\bibnamefont {Lazzeri}}, \bibinfo {author} {\bibfnamefont {L.}~\bibnamefont {Martin-Samos}}, \bibinfo {author} {\bibfnamefont {N.}~\bibnamefont {Marzari}}, \bibinfo {author} {\bibfnamefont {F.}~\bibnamefont {Mauri}}, \bibinfo {author} {\bibfnamefont {R.}~\bibnamefont {Mazzarello}}, \bibinfo {author} {\bibfnamefont {S.}~\bibnamefont {Paolini}}, \bibinfo {author} {\bibfnamefont {A.}~\bibnamefont {Pasquarello}}, \bibinfo {author} {\bibfnamefont {L.}~\bibnamefont {Paulatto}}, \bibinfo {author} {\bibfnamefont {C.}~\bibnamefont {Sbraccia}}, \bibinfo {author} {\bibfnamefont {S.}~\bibnamefont {Scandolo}}, \bibinfo {author} {\bibfnamefont {G.}~\bibnamefont {Sclauzero}}, \bibinfo {author} {\bibfnamefont {A.~P.}\ \bibnamefont {Seitsonen}}, \bibinfo {author} {\bibfnamefont {A.}~\bibnamefont {Smogunov}}, \bibinfo {author}
  {\bibfnamefont {P.}~\bibnamefont {Umari}},\ and\ \bibinfo {author} {\bibfnamefont {R.~M.}\ \bibnamefont {Wentzcovitch}},\ }\bibfield  {title} {\bibinfo {title} {Quantum espresso: a modular and open-source software project for quantum simulations of materials},\ }\href {https://doi.org/10.1088/0953-8984/21/39/395502} {\bibfield  {journal} {\bibinfo  {journal} {Journal of Physics: Condensed Matter}\ }\textbf {\bibinfo {volume} {21}},\ \bibinfo {pages} {395502} (\bibinfo {year} {2009})}\BibitemShut {NoStop}%
\bibitem [{\citenamefont {Kresse}\ and\ \citenamefont {Furthm\"uller}(1996)}]{PhysRevB.54.11169}%
  \BibitemOpen
  \bibfield  {author} {\bibinfo {author} {\bibfnamefont {G.}~\bibnamefont {Kresse}}\ and\ \bibinfo {author} {\bibfnamefont {J.}~\bibnamefont {Furthm\"uller}},\ }\bibfield  {title} {\bibinfo {title} {Efficient iterative schemes for ab initio total-energy calculations using a plane-wave basis set},\ }\href {https://doi.org/10.1103/PhysRevB.54.11169} {\bibfield  {journal} {\bibinfo  {journal} {Phys. Rev. B}\ }\textbf {\bibinfo {volume} {54}},\ \bibinfo {pages} {11169} (\bibinfo {year} {1996})}\BibitemShut {NoStop}%
\bibitem [{\citenamefont {Gonze}\ \emph {et~al.}(2020)\citenamefont {Gonze}, \citenamefont {Amadon}, \citenamefont {Antonius}, \citenamefont {Arnardi}, \citenamefont {Baguet}, \citenamefont {Beuken}, \citenamefont {Bieder}, \citenamefont {Bottin}, \citenamefont {Bouchet}, \citenamefont {Bousquet}, \citenamefont {Brouwer}, \citenamefont {Bruneval}, \citenamefont {Brunin}, \citenamefont {Cavignac}, \citenamefont {Charraud}, \citenamefont {Chen}, \citenamefont {Côté}, \citenamefont {Cottenier}, \citenamefont {Denier}, \citenamefont {Geneste}, \citenamefont {Ghosez}, \citenamefont {Giantomassi}, \citenamefont {Gillet}, \citenamefont {Gingras}, \citenamefont {Hamann}, \citenamefont {Hautier}, \citenamefont {He}, \citenamefont {Helbig}, \citenamefont {Holzwarth}, \citenamefont {Jia}, \citenamefont {Jollet}, \citenamefont {Lafargue-Dit-Hauret}, \citenamefont {Lejaeghere}, \citenamefont {Marques}, \citenamefont {Martin}, \citenamefont {Martins}, \citenamefont {Miranda}, \citenamefont {Naccarato}, \citenamefont
  {Persson}, \citenamefont {Petretto}, \citenamefont {Planes}, \citenamefont {Pouillon}, \citenamefont {Prokhorenko}, \citenamefont {Ricci}, \citenamefont {Rignanese}, \citenamefont {Romero}, \citenamefont {Schmitt}, \citenamefont {Torrent}, \citenamefont {van Setten}, \citenamefont {Troeye}, \citenamefont {Verstraete}, \citenamefont {Zérah},\ and\ \citenamefont {Zwanziger}}]{Gonze2020}%
  \BibitemOpen
  \bibfield  {author} {\bibinfo {author} {\bibfnamefont {X.}~\bibnamefont {Gonze}}, \bibinfo {author} {\bibfnamefont {B.}~\bibnamefont {Amadon}}, \bibinfo {author} {\bibfnamefont {G.}~\bibnamefont {Antonius}}, \bibinfo {author} {\bibfnamefont {F.}~\bibnamefont {Arnardi}}, \bibinfo {author} {\bibfnamefont {L.}~\bibnamefont {Baguet}}, \bibinfo {author} {\bibfnamefont {J.-M.}\ \bibnamefont {Beuken}}, \bibinfo {author} {\bibfnamefont {J.}~\bibnamefont {Bieder}}, \bibinfo {author} {\bibfnamefont {F.}~\bibnamefont {Bottin}}, \bibinfo {author} {\bibfnamefont {J.}~\bibnamefont {Bouchet}}, \bibinfo {author} {\bibfnamefont {E.}~\bibnamefont {Bousquet}}, \bibinfo {author} {\bibfnamefont {N.}~\bibnamefont {Brouwer}}, \bibinfo {author} {\bibfnamefont {F.}~\bibnamefont {Bruneval}}, \bibinfo {author} {\bibfnamefont {G.}~\bibnamefont {Brunin}}, \bibinfo {author} {\bibfnamefont {T.}~\bibnamefont {Cavignac}}, \bibinfo {author} {\bibfnamefont {J.-B.}\ \bibnamefont {Charraud}}, \bibinfo {author} {\bibfnamefont {W.}~\bibnamefont
  {Chen}}, \bibinfo {author} {\bibfnamefont {M.}~\bibnamefont {Côté}}, \bibinfo {author} {\bibfnamefont {S.}~\bibnamefont {Cottenier}}, \bibinfo {author} {\bibfnamefont {J.}~\bibnamefont {Denier}}, \bibinfo {author} {\bibfnamefont {G.}~\bibnamefont {Geneste}}, \bibinfo {author} {\bibfnamefont {P.}~\bibnamefont {Ghosez}}, \bibinfo {author} {\bibfnamefont {M.}~\bibnamefont {Giantomassi}}, \bibinfo {author} {\bibfnamefont {Y.}~\bibnamefont {Gillet}}, \bibinfo {author} {\bibfnamefont {O.}~\bibnamefont {Gingras}}, \bibinfo {author} {\bibfnamefont {D.~R.}\ \bibnamefont {Hamann}}, \bibinfo {author} {\bibfnamefont {G.}~\bibnamefont {Hautier}}, \bibinfo {author} {\bibfnamefont {X.}~\bibnamefont {He}}, \bibinfo {author} {\bibfnamefont {N.}~\bibnamefont {Helbig}}, \bibinfo {author} {\bibfnamefont {N.}~\bibnamefont {Holzwarth}}, \bibinfo {author} {\bibfnamefont {Y.}~\bibnamefont {Jia}}, \bibinfo {author} {\bibfnamefont {F.}~\bibnamefont {Jollet}}, \bibinfo {author} {\bibfnamefont {W.}~\bibnamefont
  {Lafargue-Dit-Hauret}}, \bibinfo {author} {\bibfnamefont {K.}~\bibnamefont {Lejaeghere}}, \bibinfo {author} {\bibfnamefont {M.~A.~L.}\ \bibnamefont {Marques}}, \bibinfo {author} {\bibfnamefont {A.}~\bibnamefont {Martin}}, \bibinfo {author} {\bibfnamefont {C.}~\bibnamefont {Martins}}, \bibinfo {author} {\bibfnamefont {H.~P.~C.}\ \bibnamefont {Miranda}}, \bibinfo {author} {\bibfnamefont {F.}~\bibnamefont {Naccarato}}, \bibinfo {author} {\bibfnamefont {K.}~\bibnamefont {Persson}}, \bibinfo {author} {\bibfnamefont {G.}~\bibnamefont {Petretto}}, \bibinfo {author} {\bibfnamefont {V.}~\bibnamefont {Planes}}, \bibinfo {author} {\bibfnamefont {Y.}~\bibnamefont {Pouillon}}, \bibinfo {author} {\bibfnamefont {S.}~\bibnamefont {Prokhorenko}}, \bibinfo {author} {\bibfnamefont {F.}~\bibnamefont {Ricci}}, \bibinfo {author} {\bibfnamefont {G.-M.}\ \bibnamefont {Rignanese}}, \bibinfo {author} {\bibfnamefont {A.~H.}\ \bibnamefont {Romero}}, \bibinfo {author} {\bibfnamefont {M.~M.}\ \bibnamefont {Schmitt}}, \bibinfo {author}
  {\bibfnamefont {M.}~\bibnamefont {Torrent}}, \bibinfo {author} {\bibfnamefont {M.~J.}\ \bibnamefont {van Setten}}, \bibinfo {author} {\bibfnamefont {B.~V.}\ \bibnamefont {Troeye}}, \bibinfo {author} {\bibfnamefont {M.~J.}\ \bibnamefont {Verstraete}}, \bibinfo {author} {\bibfnamefont {G.}~\bibnamefont {Zérah}},\ and\ \bibinfo {author} {\bibfnamefont {J.~W.}\ \bibnamefont {Zwanziger}},\ }\bibfield  {title} {\bibinfo {title} {The abinit project: Impact, environment and recent developments},\ }\href {https://doi.org/10.1016/j.cpc.2019.107042} {\bibfield  {journal} {\bibinfo  {journal} {Comput. Phys. Commun.}\ }\textbf {\bibinfo {volume} {248}},\ \bibinfo {pages} {107042} (\bibinfo {year} {2020})}\BibitemShut {NoStop}%
\bibitem [{\citenamefont {Hansen}\ and\ \citenamefont {McDonald}(2013)}]{HansenMcdonald}%
  \BibitemOpen
  \bibfield  {author} {\bibinfo {author} {\bibfnamefont {J.-P.}\ \bibnamefont {Hansen}}\ and\ \bibinfo {author} {\bibfnamefont {I.~R.}\ \bibnamefont {McDonald}},\ }\href@noop {} {\emph {\bibinfo {title} {Theory of simple liquids: with applications to soft matter}}}\ (\bibinfo  {publisher} {Academic press},\ \bibinfo {year} {2013})\BibitemShut {NoStop}%
\bibitem [{\citenamefont {Shaffer}\ and\ \citenamefont {Starrett}(2020)}]{PhysRevE.101.013208}%
  \BibitemOpen
  \bibfield  {author} {\bibinfo {author} {\bibfnamefont {N.~R.}\ \bibnamefont {Shaffer}}\ and\ \bibinfo {author} {\bibfnamefont {C.~E.}\ \bibnamefont {Starrett}},\ }\bibfield  {title} {\bibinfo {title} {Correlations between conduction electrons in dense plasmas},\ }\href {https://doi.org/10.1103/PhysRevE.101.013208} {\bibfield  {journal} {\bibinfo  {journal} {Phys. Rev. E}\ }\textbf {\bibinfo {volume} {101}},\ \bibinfo {pages} {013208} (\bibinfo {year} {2020})}\BibitemShut {NoStop}%
\bibitem [{\citenamefont {Rosenfeld}\ and\ \citenamefont {Ashcroft}(1979)}]{rosenfeld1979theory}%
  \BibitemOpen
  \bibfield  {author} {\bibinfo {author} {\bibfnamefont {Y.}~\bibnamefont {Rosenfeld}}\ and\ \bibinfo {author} {\bibfnamefont {N.}~\bibnamefont {Ashcroft}},\ }\bibfield  {title} {\bibinfo {title} {Theory of simple classical fluids: Universality in the short-range structure},\ }\href@noop {} {\bibfield  {journal} {\bibinfo  {journal} {Physical Review A}\ }\textbf {\bibinfo {volume} {20}},\ \bibinfo {pages} {1208} (\bibinfo {year} {1979})}\BibitemShut {NoStop}%
\bibitem [{\citenamefont {Iyetomi}\ \emph {et~al.}(1992)\citenamefont {Iyetomi}, \citenamefont {Ogata},\ and\ \citenamefont {Ichimaru}}]{PhysRevA.46.1051}%
  \BibitemOpen
  \bibfield  {author} {\bibinfo {author} {\bibfnamefont {H.}~\bibnamefont {Iyetomi}}, \bibinfo {author} {\bibfnamefont {S.}~\bibnamefont {Ogata}},\ and\ \bibinfo {author} {\bibfnamefont {S.}~\bibnamefont {Ichimaru}},\ }\bibfield  {title} {\bibinfo {title} {Bridge functions and improvement on the hypernetted-chain approximation for classical one-component plasmas},\ }\href {https://doi.org/10.1103/PhysRevA.46.1051} {\bibfield  {journal} {\bibinfo  {journal} {Phys. Rev. A}\ }\textbf {\bibinfo {volume} {46}},\ \bibinfo {pages} {1051} (\bibinfo {year} {1992})}\BibitemShut {NoStop}%
\bibitem [{\citenamefont {Daughton}\ \emph {et~al.}(2000)\citenamefont {Daughton}, \citenamefont {Murillo},\ and\ \citenamefont {Thode}}]{PhysRevE.61.2129}%
  \BibitemOpen
  \bibfield  {author} {\bibinfo {author} {\bibfnamefont {W.}~\bibnamefont {Daughton}}, \bibinfo {author} {\bibfnamefont {M.~S.}\ \bibnamefont {Murillo}},\ and\ \bibinfo {author} {\bibfnamefont {L.}~\bibnamefont {Thode}},\ }\bibfield  {title} {\bibinfo {title} {Empirical bridge function for strongly coupled yukawa systems},\ }\href {https://doi.org/10.1103/PhysRevE.61.2129} {\bibfield  {journal} {\bibinfo  {journal} {Phys. Rev. E}\ }\textbf {\bibinfo {volume} {61}},\ \bibinfo {pages} {2129} (\bibinfo {year} {2000})}\BibitemShut {NoStop}%
\bibitem [{\citenamefont {Shaffer}\ \emph {et~al.}(2017)\citenamefont {Shaffer}, \citenamefont {Tiwari},\ and\ \citenamefont {Baalrud}}]{Shaffer2017}%
  \BibitemOpen
  \bibfield  {author} {\bibinfo {author} {\bibfnamefont {N.~R.}\ \bibnamefont {Shaffer}}, \bibinfo {author} {\bibfnamefont {S.~K.}\ \bibnamefont {Tiwari}},\ and\ \bibinfo {author} {\bibfnamefont {S.~D.}\ \bibnamefont {Baalrud}},\ }\bibfield  {title} {\bibinfo {title} {Pair correlation functions of strongly coupled twotemperature plasma},\ }\href
  {https://watermark.silverchair.com/092703_1_online.pdf?token=AQECAHi208BE49Ooan9kkhW_Ercy7Dm3ZL_9Cf3qfKAc485ysgAABTAwggUsBgkqhkiG9w0BBwagggUdMIIFGQIBADCCBRIGCSqGSIb3DQEHATAeBglghkgBZQMEAS4wEQQM9tZDWtbveVWScqsPAgEQgIIE4ywHFotojwrsN3tvTZFBo_o7tZpjIGaXKvVysJyZOlap5vHFInmJ163e1zG3iv9n-L8Wb-JBIN5QxwfXfS3GVNJIR1mUBc0gBQe2UMzlvUrB-s2-zJmKCE7WcqeDGsKZcpmBOOs3F1nA441xlJQzLNhLsKqOoD4EdaLMpeKoGXDPzxaZ0bFiYfkby-iReUfTlECuEEkLvkVc6Zs7Rqwy_N45p0ZmUqZ09G0uZiQcFWKRJXQdIbC8nhoMjFc9wLk6M2XM8D_iSlBADFv9BoAn3_eaml_5TzFdlrdTiJ684YbcyLbiRNtWQgCc81pLdyyZy60gcV1D4rXoKZjmY_4fwiw10nJ38Ln0-fOdKcWIJ6Eo2s4HXnSRnmW5PRyy13ZO2OyDOvZpa8ojv4woWV6JXQhgjSjZtgVzuTLu6MWUNL3sVBwWUKakTw-YVesRobNDIlGtuNn4hpTt4Q4TTLhOUafDDFxLr7pzPCuqFw_w-1LkUmSbPlB4HX39CfL0Wys8Oyc6JMfC11eh-fyosDa_NagFdqZvwrSpJT9o4rbzlgcaiZNy4dMsxLmrVvtYDh9h3RzsQqddcGllvpBPy7-qxCAnb2x61ktz3TMWUbvl0hzwlfYPZRQXr-j03WPyvQNk7l6nP3WDk6gRaFqW7LtxrU_5mVJo5AvBkAFcmsafwlGJNZq0uv2KYZieC0SgoVUmFyYbdfY6pDQRHg5IkMxJT2YAWU0S9_7jHtJ6t0ZYytLk4wVUxvmkhVdWJXr5PG5Hn2H1aYPfeXCxwjwTGxSWaxeVBa4AavP8NLsOc1BPzRIyNsnXNkDclPKQWOf2TTwLtNbiKfuOA71E09uiOed265LJLP0WeYeYekYB57sC78BszdzmIVWDosGvDGsyUS4XLvSrybrk3mFZMRlPm2JZ-DMl9fCxDsqaR1Pbkq22vPABeIs3Sazyy2Atucb3hCtkYwIIbQxgn5RS5a9YlkRdu13knZ9QCsKoE-PLvPNEoy3_CMmFGjYsUss0MyOhqLSHHfeE9pCwlJSH4Kg1r2o4O3RAriJDcuzmKGvRbaQeEzyyey-pUTVyQWVlqOyey3Tkqs-mgAPLGecmPETsCILQ20m51is5tia9hFVuq0UoxfugwMyVOzddInftg84uj7BztAlpuKe5sZkMcNUauyW5SCSpUxx3SujyARqhcQ51aCRZEYcN_H0YKa6GsqHRbTxUItfZ9Jd7exGciBPCJEvGrwi5JfIdNrE2hKJr3iWY-AMAJhZfY3MBMR1qGLW7MQxPFs3JN4_73zsBAM3DtBesb89shLifZiMC6-hrPc7DBUw_bCZyKhDMPrn7vKQ8DMu4cZvDtnd7hMd4ubrp7mH3PBaVgxdsmR6jt6NMhctduKZyv3FsLAHLidnugRBPpiDfxz-3Nd95-kDmttEJmA5-1LKdOS3VydNEwXIfPSDVg4VQBYu0b_OhV2glJSr4LEvIcEyBDMkZoHE7VKmfPXZHRBaU9nVsZz_5VahApzEBnA7MDARKAWhf_OiBvkaLpY-6GPtK27PoUcerDTUylLmpOl4J-rnNoVtrtQShGQqyYhd8mMiCBRgzmtyhk1cQ0IkoLp4uxrYQFhaCIrBI3Ajy6VwxMe8B4Ln5JHnNGX6wozvsLw8L9OD5s-kqFloYaof-zqjjTg}
  {\bibfield  {journal} {\bibinfo  {journal} {Physics of Plasmas 24, 092703 (2017)}\ } (\bibinfo {year} {2017})}\BibitemShut {NoStop}%
\bibitem [{\citenamefont {Seuferling}\ \emph {et~al.}(1989)\citenamefont {Seuferling}, \citenamefont {Vogel},\ and\ \citenamefont {Toepffer}}]{SVT1989}%
  \BibitemOpen
  \bibfield  {author} {\bibinfo {author} {\bibfnamefont {P.}~\bibnamefont {Seuferling}}, \bibinfo {author} {\bibfnamefont {J.}~\bibnamefont {Vogel}},\ and\ \bibinfo {author} {\bibfnamefont {C.}~\bibnamefont {Toepffer}},\ }\bibfield  {title} {\bibinfo {title} {Correlations in a two-temperature plasma},\ }\href {https://doi.org/10.1103/PhysRevA.40.323} {\bibfield  {journal} {\bibinfo  {journal} {Phys. Rev. A}\ }\textbf {\bibinfo {volume} {40}},\ \bibinfo {pages} {323} (\bibinfo {year} {1989})}\BibitemShut {NoStop}%
\bibitem [{\citenamefont {Anta}\ and\ \citenamefont {Louis}(2000)}]{Anta2000}%
  \BibitemOpen
  \bibfield  {author} {\bibinfo {author} {\bibfnamefont {J.}~\bibnamefont {Anta}}\ and\ \bibinfo {author} {\bibfnamefont {A.}~\bibnamefont {Louis}},\ }\bibfield  {title} {\bibinfo {title} {Probing ion-ion and electron-ion correlations in liquid metals within the quantum hypernetted chain approximation},\ }\href {https://doi.org/10.1103/PhysRevB.61.11400} {\bibfield  {journal} {\bibinfo  {journal} {Physical Review B}\ }\textbf {\bibinfo {volume} {61}},\ \bibinfo {pages} {11400} (\bibinfo {year} {2000})}\BibitemShut {NoStop}%
\bibitem [{\citenamefont {Chihara}(1978)}]{Chihara1978}%
  \BibitemOpen
  \bibfield  {author} {\bibinfo {author} {\bibfnamefont {J.}~\bibnamefont {Chihara}},\ }\bibfield  {title} {\bibinfo {title} {Derivation of quantal hyper-netted chain equation from the kohn-sham theory},\ }\href {https://doi.org/10.1143/PTP.59.76} {\bibfield  {journal} {\bibinfo  {journal} {Progress of Theoretical Physics}\ }\textbf {\bibinfo {volume} {59}},\ \bibinfo {pages} {76} (\bibinfo {year} {1978})}\BibitemShut {NoStop}%
\bibitem [{\citenamefont {Chihara}(1985)}]{Chihara_1985}%
  \BibitemOpen
  \bibfield  {author} {\bibinfo {author} {\bibfnamefont {J.}~\bibnamefont {Chihara}},\ }\bibfield  {title} {\bibinfo {title} {Liquid metals and plasmas as nucleus-electron mixtures},\ }\href {https://doi.org/10.1088/0022-3719/18/16/008} {\bibfield  {journal} {\bibinfo  {journal} {Journal of Physics C: Solid State Physics}\ }\textbf {\bibinfo {volume} {18}},\ \bibinfo {pages} {3103} (\bibinfo {year} {1985})}\BibitemShut {NoStop}%
\bibitem [{\citenamefont {Kresse}\ and\ \citenamefont {Hafner}(1993)}]{PhysRevB.47.558}%
  \BibitemOpen
  \bibfield  {author} {\bibinfo {author} {\bibfnamefont {G.}~\bibnamefont {Kresse}}\ and\ \bibinfo {author} {\bibfnamefont {J.}~\bibnamefont {Hafner}},\ }\bibfield  {title} {\bibinfo {title} {Ab initio molecular dynamics for liquid metals},\ }\href {https://doi.org/10.1103/PhysRevB.47.558} {\bibfield  {journal} {\bibinfo  {journal} {Phys. Rev. B}\ }\textbf {\bibinfo {volume} {47}},\ \bibinfo {pages} {558} (\bibinfo {year} {1993})}\BibitemShut {NoStop}%
\bibitem [{\citenamefont {Kresse}\ and\ \citenamefont {Furthmüller}(1996)}]{KRESSE199615}%
  \BibitemOpen
  \bibfield  {author} {\bibinfo {author} {\bibfnamefont {G.}~\bibnamefont {Kresse}}\ and\ \bibinfo {author} {\bibfnamefont {J.}~\bibnamefont {Furthmüller}},\ }\bibfield  {title} {\bibinfo {title} {Efficiency of ab-initio total energy calculations for metals and semiconductors using a plane-wave basis set},\ }\href {https://doi.org/https://doi.org/10.1016/0927-0256(96)00008-0} {\bibfield  {journal} {\bibinfo  {journal} {Computational Materials Science}\ }\textbf {\bibinfo {volume} {6}},\ \bibinfo {pages} {15} (\bibinfo {year} {1996})}\BibitemShut {NoStop}%
\bibitem [{\citenamefont {Kresse}\ and\ \citenamefont {Joubert}(1999)}]{PhysRevB.59.1758}%
  \BibitemOpen
  \bibfield  {author} {\bibinfo {author} {\bibfnamefont {G.}~\bibnamefont {Kresse}}\ and\ \bibinfo {author} {\bibfnamefont {D.}~\bibnamefont {Joubert}},\ }\bibfield  {title} {\bibinfo {title} {From ultrasoft pseudopotentials to the projector augmented-wave method},\ }\href {https://doi.org/10.1103/PhysRevB.59.1758} {\bibfield  {journal} {\bibinfo  {journal} {Phys. Rev. B}\ }\textbf {\bibinfo {volume} {59}},\ \bibinfo {pages} {1758} (\bibinfo {year} {1999})}\BibitemShut {NoStop}%
\bibitem [{\citenamefont {Baldereschi}(1973)}]{PhysRevB.7.5212}%
  \BibitemOpen
  \bibfield  {author} {\bibinfo {author} {\bibfnamefont {A.}~\bibnamefont {Baldereschi}},\ }\bibfield  {title} {\bibinfo {title} {Mean-value point in the brillouin zone},\ }\href {https://doi.org/10.1103/PhysRevB.7.5212} {\bibfield  {journal} {\bibinfo  {journal} {Phys. Rev. B}\ }\textbf {\bibinfo {volume} {7}},\ \bibinfo {pages} {5212} (\bibinfo {year} {1973})}\BibitemShut {NoStop}%
\bibitem [{\citenamefont {Thompson}\ \emph {et~al.}(2022)\citenamefont {Thompson}, \citenamefont {Aktulga}, \citenamefont {Berger}, \citenamefont {Bolintineanu}, \citenamefont {Brown}, \citenamefont {Crozier}, \citenamefont {{in 't Veld}}, \citenamefont {Kohlmeyer}, \citenamefont {Moore}, \citenamefont {Nguyen}, \citenamefont {Shan}, \citenamefont {Stevens}, \citenamefont {Tranchida}, \citenamefont {Trott},\ and\ \citenamefont {Plimpton}}]{THOMPSON2022108171}%
  \BibitemOpen
  \bibfield  {author} {\bibinfo {author} {\bibfnamefont {A.~P.}\ \bibnamefont {Thompson}}, \bibinfo {author} {\bibfnamefont {H.~M.}\ \bibnamefont {Aktulga}}, \bibinfo {author} {\bibfnamefont {R.}~\bibnamefont {Berger}}, \bibinfo {author} {\bibfnamefont {D.~S.}\ \bibnamefont {Bolintineanu}}, \bibinfo {author} {\bibfnamefont {W.~M.}\ \bibnamefont {Brown}}, \bibinfo {author} {\bibfnamefont {P.~S.}\ \bibnamefont {Crozier}}, \bibinfo {author} {\bibfnamefont {P.~J.}\ \bibnamefont {{in 't Veld}}}, \bibinfo {author} {\bibfnamefont {A.}~\bibnamefont {Kohlmeyer}}, \bibinfo {author} {\bibfnamefont {S.~G.}\ \bibnamefont {Moore}}, \bibinfo {author} {\bibfnamefont {T.~D.}\ \bibnamefont {Nguyen}}, \bibinfo {author} {\bibfnamefont {R.}~\bibnamefont {Shan}}, \bibinfo {author} {\bibfnamefont {M.~J.}\ \bibnamefont {Stevens}}, \bibinfo {author} {\bibfnamefont {J.}~\bibnamefont {Tranchida}}, \bibinfo {author} {\bibfnamefont {C.}~\bibnamefont {Trott}},\ and\ \bibinfo {author} {\bibfnamefont {S.~J.}\ \bibnamefont {Plimpton}},\
  }\bibfield  {title} {\bibinfo {title} {Lammps - a flexible simulation tool for particle-based materials modeling at the atomic, meso, and continuum scales},\ }\href {https://doi.org/https://doi.org/10.1016/j.cpc.2021.108171} {\bibfield  {journal} {\bibinfo  {journal} {Computer Physics Communications}\ }\textbf {\bibinfo {volume} {271}},\ \bibinfo {pages} {108171} (\bibinfo {year} {2022})}\BibitemShut {NoStop}%
\bibitem [{\citenamefont {More}(1985)}]{more1985pressure}%
  \BibitemOpen
  \bibfield  {author} {\bibinfo {author} {\bibfnamefont {R.}~\bibnamefont {More}},\ }\bibfield  {title} {\bibinfo {title} {Pressure ionization, resonances, and the continuity of bound and free states},\ }in\ \href@noop {} {\emph {\bibinfo {booktitle} {Advances in atomic and molecular physics}}},\ Vol.~\bibinfo {volume} {21}\ (\bibinfo  {publisher} {Elsevier},\ \bibinfo {year} {1985})\ pp.\ \bibinfo {pages} {305--356}\BibitemShut {NoStop}%
\bibitem [{\citenamefont {Daligault}\ \emph {et~al.}(2016)\citenamefont {Daligault}, \citenamefont {Baalrud}, \citenamefont {Starrett}, \citenamefont {Saumon},\ and\ \citenamefont {Sjostrom}}]{PhysRevLett.116.075002}%
  \BibitemOpen
  \bibfield  {author} {\bibinfo {author} {\bibfnamefont {J.}~\bibnamefont {Daligault}}, \bibinfo {author} {\bibfnamefont {S.~D.}\ \bibnamefont {Baalrud}}, \bibinfo {author} {\bibfnamefont {C.~E.}\ \bibnamefont {Starrett}}, \bibinfo {author} {\bibfnamefont {D.}~\bibnamefont {Saumon}},\ and\ \bibinfo {author} {\bibfnamefont {T.}~\bibnamefont {Sjostrom}},\ }\bibfield  {title} {\bibinfo {title} {Ionic transport coefficients of dense plasmas without molecular dynamics},\ }\href {https://doi.org/10.1103/PhysRevLett.116.075002} {\bibfield  {journal} {\bibinfo  {journal} {Phys. Rev. Lett.}\ }\textbf {\bibinfo {volume} {116}},\ \bibinfo {pages} {075002} (\bibinfo {year} {2016})}\BibitemShut {NoStop}%
\bibitem [{\citenamefont {Giannozzi}\ \emph {et~al.}(2017)\citenamefont {Giannozzi}, \citenamefont {Andreussi}, \citenamefont {Brumme}, \citenamefont {Bunau}, \citenamefont {Nardelli}, \citenamefont {Calandra}, \citenamefont {Car}, \citenamefont {Cavazzoni}, \citenamefont {Ceresoli}, \citenamefont {Cococcioni}, \citenamefont {Colonna}, \citenamefont {Carnimeo}, \citenamefont {Corso}, \citenamefont {de~Gironcoli}, \citenamefont {Delugas}, \citenamefont {DiStasio}, \citenamefont {Ferretti}, \citenamefont {Floris}, \citenamefont {Fratesi}, \citenamefont {Fugallo}, \citenamefont {Gebauer}, \citenamefont {Gerstmann}, \citenamefont {Giustino}, \citenamefont {Gorni}, \citenamefont {Jia}, \citenamefont {Kawamura}, \citenamefont {Ko}, \citenamefont {Kokalj}, \citenamefont {Küçükbenli}, \citenamefont {Lazzeri}, \citenamefont {Marsili}, \citenamefont {Marzari}, \citenamefont {Mauri}, \citenamefont {Nguyen}, \citenamefont {Nguyen}, \citenamefont {de-la Roza}, \citenamefont {Paulatto}, \citenamefont {Poncé},
  \citenamefont {Rocca}, \citenamefont {Sabatini}, \citenamefont {Santra}, \citenamefont {Schlipf}, \citenamefont {Seitsonen}, \citenamefont {Smogunov}, \citenamefont {Timrov}, \citenamefont {Thonhauser}, \citenamefont {Umari}, \citenamefont {Vast}, \citenamefont {Wu},\ and\ \citenamefont {Baroni}}]{Giannozzi_2017}%
  \BibitemOpen
  \bibfield  {author} {\bibinfo {author} {\bibfnamefont {P.}~\bibnamefont {Giannozzi}}, \bibinfo {author} {\bibfnamefont {O.}~\bibnamefont {Andreussi}}, \bibinfo {author} {\bibfnamefont {T.}~\bibnamefont {Brumme}}, \bibinfo {author} {\bibfnamefont {O.}~\bibnamefont {Bunau}}, \bibinfo {author} {\bibfnamefont {M.~B.}\ \bibnamefont {Nardelli}}, \bibinfo {author} {\bibfnamefont {M.}~\bibnamefont {Calandra}}, \bibinfo {author} {\bibfnamefont {R.}~\bibnamefont {Car}}, \bibinfo {author} {\bibfnamefont {C.}~\bibnamefont {Cavazzoni}}, \bibinfo {author} {\bibfnamefont {D.}~\bibnamefont {Ceresoli}}, \bibinfo {author} {\bibfnamefont {M.}~\bibnamefont {Cococcioni}}, \bibinfo {author} {\bibfnamefont {N.}~\bibnamefont {Colonna}}, \bibinfo {author} {\bibfnamefont {I.}~\bibnamefont {Carnimeo}}, \bibinfo {author} {\bibfnamefont {A.~D.}\ \bibnamefont {Corso}}, \bibinfo {author} {\bibfnamefont {S.}~\bibnamefont {de~Gironcoli}}, \bibinfo {author} {\bibfnamefont {P.}~\bibnamefont {Delugas}}, \bibinfo {author} {\bibfnamefont
  {R.~A.}\ \bibnamefont {DiStasio}}, \bibinfo {author} {\bibfnamefont {A.}~\bibnamefont {Ferretti}}, \bibinfo {author} {\bibfnamefont {A.}~\bibnamefont {Floris}}, \bibinfo {author} {\bibfnamefont {G.}~\bibnamefont {Fratesi}}, \bibinfo {author} {\bibfnamefont {G.}~\bibnamefont {Fugallo}}, \bibinfo {author} {\bibfnamefont {R.}~\bibnamefont {Gebauer}}, \bibinfo {author} {\bibfnamefont {U.}~\bibnamefont {Gerstmann}}, \bibinfo {author} {\bibfnamefont {F.}~\bibnamefont {Giustino}}, \bibinfo {author} {\bibfnamefont {T.}~\bibnamefont {Gorni}}, \bibinfo {author} {\bibfnamefont {J.}~\bibnamefont {Jia}}, \bibinfo {author} {\bibfnamefont {M.}~\bibnamefont {Kawamura}}, \bibinfo {author} {\bibfnamefont {H.-Y.}\ \bibnamefont {Ko}}, \bibinfo {author} {\bibfnamefont {A.}~\bibnamefont {Kokalj}}, \bibinfo {author} {\bibfnamefont {E.}~\bibnamefont {Küçükbenli}}, \bibinfo {author} {\bibfnamefont {M.}~\bibnamefont {Lazzeri}}, \bibinfo {author} {\bibfnamefont {M.}~\bibnamefont {Marsili}}, \bibinfo {author} {\bibfnamefont
  {N.}~\bibnamefont {Marzari}}, \bibinfo {author} {\bibfnamefont {F.}~\bibnamefont {Mauri}}, \bibinfo {author} {\bibfnamefont {N.~L.}\ \bibnamefont {Nguyen}}, \bibinfo {author} {\bibfnamefont {H.-V.}\ \bibnamefont {Nguyen}}, \bibinfo {author} {\bibfnamefont {A.~O.}\ \bibnamefont {de-la Roza}}, \bibinfo {author} {\bibfnamefont {L.}~\bibnamefont {Paulatto}}, \bibinfo {author} {\bibfnamefont {S.}~\bibnamefont {Poncé}}, \bibinfo {author} {\bibfnamefont {D.}~\bibnamefont {Rocca}}, \bibinfo {author} {\bibfnamefont {R.}~\bibnamefont {Sabatini}}, \bibinfo {author} {\bibfnamefont {B.}~\bibnamefont {Santra}}, \bibinfo {author} {\bibfnamefont {M.}~\bibnamefont {Schlipf}}, \bibinfo {author} {\bibfnamefont {A.~P.}\ \bibnamefont {Seitsonen}}, \bibinfo {author} {\bibfnamefont {A.}~\bibnamefont {Smogunov}}, \bibinfo {author} {\bibfnamefont {I.}~\bibnamefont {Timrov}}, \bibinfo {author} {\bibfnamefont {T.}~\bibnamefont {Thonhauser}}, \bibinfo {author} {\bibfnamefont {P.}~\bibnamefont {Umari}}, \bibinfo {author}
  {\bibfnamefont {N.}~\bibnamefont {Vast}}, \bibinfo {author} {\bibfnamefont {X.}~\bibnamefont {Wu}},\ and\ \bibinfo {author} {\bibfnamefont {S.}~\bibnamefont {Baroni}},\ }\bibfield  {title} {\bibinfo {title} {Advanced capabilities for materials modelling with quantum espresso},\ }\href {https://doi.org/10.1088/1361-648X/aa8f79} {\bibfield  {journal} {\bibinfo  {journal} {Journal of Physics: Condensed Matter}\ }\textbf {\bibinfo {volume} {29}},\ \bibinfo {pages} {465901} (\bibinfo {year} {2017})}\BibitemShut {NoStop}%
\bibitem [{\citenamefont {Hou}\ and\ \citenamefont {Yuan}(2009)}]{PhysRevE.79.016402}%
  \BibitemOpen
  \bibfield  {author} {\bibinfo {author} {\bibfnamefont {Y.}~\bibnamefont {Hou}}\ and\ \bibinfo {author} {\bibfnamefont {J.}~\bibnamefont {Yuan}},\ }\bibfield  {title} {\bibinfo {title} {Alternative ion-ion pair-potential model applied to molecular dynamics simulations of hot and dense plasmas: Al and fe as examples},\ }\href {https://doi.org/10.1103/PhysRevE.79.016402} {\bibfield  {journal} {\bibinfo  {journal} {Phys. Rev. E}\ }\textbf {\bibinfo {volume} {79}},\ \bibinfo {pages} {016402} (\bibinfo {year} {2009})}\BibitemShut {NoStop}%
\bibitem [{\citenamefont {Drake}\ and\ \citenamefont {Drake}(2006)}]{drake2006introduction}%
  \BibitemOpen
  \bibfield  {author} {\bibinfo {author} {\bibfnamefont {R.~P.}\ \bibnamefont {Drake}}\ and\ \bibinfo {author} {\bibfnamefont {R.~P.}\ \bibnamefont {Drake}},\ }\href@noop {} {\emph {\bibinfo {title} {Introduction to high-energy-density physics}}}\ (\bibinfo  {publisher} {Springer},\ \bibinfo {year} {2006})\BibitemShut {NoStop}%
\bibitem [{\citenamefont {Chapman}\ and\ \citenamefont {Cowling}(1991)}]{Chapman1991-as}%
  \BibitemOpen
  \bibfield  {author} {\bibinfo {author} {\bibfnamefont {S.}~\bibnamefont {Chapman}}\ and\ \bibinfo {author} {\bibfnamefont {T.~G.}\ \bibnamefont {Cowling}},\ }\href@noop {} {\emph {\bibinfo {title} {Cambridge mathematical library: The mathematical theory of non-uniform gases: An account of the kinetic theory of viscosity, thermal conduction and diffusion in gases}}}\ (\bibinfo  {publisher} {Cambridge University Press},\ \bibinfo {address} {Cambridge, England},\ \bibinfo {year} {1991})\BibitemShut {NoStop}%
\bibitem [{\citenamefont {Dharma-wardana}\ and\ \citenamefont {Murillo}(2008)}]{PhysRevE.77.026401}%
  \BibitemOpen
  \bibfield  {author} {\bibinfo {author} {\bibfnamefont {M.~W.~C.}\ \bibnamefont {Dharma-wardana}}\ and\ \bibinfo {author} {\bibfnamefont {M.~S.}\ \bibnamefont {Murillo}},\ }\bibfield  {title} {\bibinfo {title} {Pair-distribution functions of two-temperature two-mass systems: Comparison of molecular dynamics, classical-map hypernetted chain, quantum monte carlo, and kohn-sham calculations for dense hydrogen},\ }\href {https://doi.org/10.1103/PhysRevE.77.026401} {\bibfield  {journal} {\bibinfo  {journal} {Phys. Rev. E}\ }\textbf {\bibinfo {volume} {77}},\ \bibinfo {pages} {026401} (\bibinfo {year} {2008})}\BibitemShut {NoStop}%
\bibitem [{\citenamefont {Witte}\ \emph {et~al.}(2019)\citenamefont {Witte}, \citenamefont {R\"opke}, \citenamefont {Neumayer}, \citenamefont {French}, \citenamefont {Sperling}, \citenamefont {Recoules}, \citenamefont {Glenzer},\ and\ \citenamefont {Redmer}}]{DWcomment2019}%
  \BibitemOpen
  \bibfield  {author} {\bibinfo {author} {\bibfnamefont {B.~B.~L.}\ \bibnamefont {Witte}}, \bibinfo {author} {\bibfnamefont {G.}~\bibnamefont {R\"opke}}, \bibinfo {author} {\bibfnamefont {P.}~\bibnamefont {Neumayer}}, \bibinfo {author} {\bibfnamefont {M.}~\bibnamefont {French}}, \bibinfo {author} {\bibfnamefont {P.}~\bibnamefont {Sperling}}, \bibinfo {author} {\bibfnamefont {V.}~\bibnamefont {Recoules}}, \bibinfo {author} {\bibfnamefont {S.~H.}\ \bibnamefont {Glenzer}},\ and\ \bibinfo {author} {\bibfnamefont {R.}~\bibnamefont {Redmer}},\ }\bibfield  {title} {\bibinfo {title} {Comment on ``isochoric, isobaric, and ultrafast conductivities of aluminum, lithium, and carbon in the warm dense matter regime''},\ }\href {https://doi.org/10.1103/PhysRevE.99.047201} {\bibfield  {journal} {\bibinfo  {journal} {Phys. Rev. E}\ }\textbf {\bibinfo {volume} {99}},\ \bibinfo {pages} {047201} (\bibinfo {year} {2019})}\BibitemShut {NoStop}%
\bibitem [{\citenamefont {Fetsch}\ \emph {et~al.}(2023)\citenamefont {Fetsch}, \citenamefont {Foster},\ and\ \citenamefont {Fisch}}]{Fetsch2023}%
  \BibitemOpen
  \bibfield  {author} {\bibinfo {author} {\bibfnamefont {H.}~\bibnamefont {Fetsch}}, \bibinfo {author} {\bibfnamefont {T.~E.}\ \bibnamefont {Foster}},\ and\ \bibinfo {author} {\bibfnamefont {N.~J.}\ \bibnamefont {Fisch}},\ }\bibfield  {title} {\bibinfo {title} {Temperature separation under compression of moderately coupled plasma},\ }\href {https://doi.org/10.1017/S0022377823000776} {\bibfield  {journal} {\bibinfo  {journal} {Journal of Plasma Physics}\ }\textbf {\bibinfo {volume} {89}},\ \bibinfo {pages} {905890510} (\bibinfo {year} {2023})}\BibitemShut {NoStop}%
\bibitem [{\citenamefont {Kardar}(2007)}]{Kardar_2007}%
  \BibitemOpen
  \bibfield  {author} {\bibinfo {author} {\bibfnamefont {M.}~\bibnamefont {Kardar}},\ }\href@noop {} {\emph {\bibinfo {title} {Statistical Physics of Particles}}}\ (\bibinfo  {publisher} {Cambridge University Press},\ \bibinfo {year} {2007})\BibitemShut {NoStop}%
\end{thebibliography}%

\appendix
\newpage
\section{The Two-Temperature Partition Function in the Born-Oppenheimer Approximation   }
\label{app:thermo derivation}

Here we derive the partition function for a two-temperature system of electrons and ions, under the assumption that the electrons are inertia-less and adiabatically move to minimize their thermodynamic potential. We will arrive at an expression that is highly similar to \cite{BoerckerMore1986}. 

\begin{figure}
    \centering
    \includegraphics[width=0.6\linewidth]{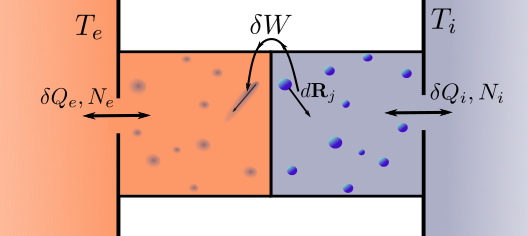}
    \caption{Two heat reservoirs at $T_e$ and $T_I$ exchange heat with the electron and ion subsystems, respectively. Movement of individual ions (right) perturbs the electron subsystem (left), causing energy transfer through work $\delta W$.   }
    \label{fig:statistical_mechanics_visualization}
\end{figure}

We consider separate electron and ion subsystems, each attached to their own heat reservoir, depicted in Fig.~\ref{fig:statistical_mechanics_visualization}. We imagine each reservoir is in the microcanonical ensemble, with the electron reservoir at temperature $T_e$, and the ion reservoir at $T_I$. Energy exchange between subsystems is allowed via work.

For the electron subsystem, the inertialess assumption implies we treat the ions as an external potential $V_{\rm ext}$. Thus, the statistical mechanics of the electrons is that of the grand canonical ensemble with an external potential
\begin{align}
\label{eq:Ωe_Vext}
    d\Omega_e = -S_e dT_e - N_e d\mu_e + \int d^3 \r n_e \delta V_{\rm ext}.
\end{align}
This is what is assumed in Born-Oppenheimer KSDFT, and the Euler equation for this grand potential is solved for each ion MD timestep. If the electrons are in a different ensemble, the derivation can be easily adapted. 
The difficulty in deriving the corresponding ion statistical mechanics lies in defining precisely the effect of this electronic subsytem on the ionic one. In \cite{Fetsch2023} a force balancing argument is made to derive the energy exchange of the electronic and ionic subsystems. Motivated by this, we proceed with a similar picture but in terms of ion subsystem microstates and energy exchange. 
First, we convert the dependence on the external potential, to an explicit dependence on the ion coordinates,
\begin{align}
\label{eq:dΩe}
d\Omega_e = -S_e dT_e - N_e d\mu_e - \sum_j {\vb {F}}_j \cdot d \vb R_j,
\end{align}
for a force-like $\vb F_j$ acting on each ion $j$ from the,
\begin{align}
    \vb F_j = -\int d^3 {\vb r} n_e(\vb r) {\vb \nabla_{R_j}} V_{eI} ({\vb r})
\end{align}
Assume the ions are described by a subsystem $\mathcal{S}_i$ attached to a much larger reservoir $\mathcal{R}_i$ in the normal thermodynamic sense\cite{Kardar_2007}. The reservoir is in the microcanonical ensemble under the opposite force from the ion subsystem, means, as usual, that it mimics the force from the electronic subsystem,
\begin{align}
    dU^{\mathcal{R}_I} = T_I dS^{\mathcal{R}_I} + \mu_I dN_I^{\mathcal{R}_I} - \sum_j {\vb {F}}_j \cdot d \vb R_j
\end{align}
Assume the ion subsystem is in a specific reference microstate $\mu_I^0$ corresponding to a specific point in classical phase space, $\vb q_I^0$, with energy $E_I^0$ determined by the ionic Hamiltonian $H_{II} = K_{II} + U_{II}$. 

The probability of this state is determined by the number of reservoir microstates, or equivalently in terms of entropy of the bath when the microstate is in state $\mu_I^0$, $S_{\mathcal{R}_j}(\mu_I^0)$ as\cite{Kardar_2007} 
\begin{align}
\label{eq:microstate_probability}
    p(\mu^0_I) \propto e^{S_{\mathcal{R}_I}(\mu_I^0)}. 
\end{align}
Consider then a small displacement in phase space with fixed momenta and particle number. The subsystem energy changes by $ dV_{II}$. The entropy of the bath changes by 
\begin{align}
      dS^{\mathcal{R}_I} = -\beta_I d V_{II}  + \beta_I \sum_j {\vb {F}}_j \cdot d \vb R_j
\end{align}
From Eq.~\eqref{eq:microstate_probability} and Eq.~\eqref{eq:dΩe}, we have a probability of 
\begin{align}
        p(\mu^\delta_I) &= p(\mu^0_I) \exp\left[-\beta_I \sum_j \left( \pdv{V_{II}}{\vb R_j}  + \pdv{\Omega_e}{\vb R_j}\Big|_{T_e, \mu_e} \right) \cdot d \vb R_j \right]
\end{align}
We now know the probability ratio between any two points. Multiplying successive neighboring points in phase space along any path allows us to get the probability of a generic point in phase space in the form of an integral over this path. We parameterize it with $\lambda$ such that at $\lambda=0$ we have state $\mu^0_I$ and $\{\vb R_I^0\}$, and at $\lambda=1$ we have state $\mu^1_I$ and $\{\vb R_I^1\}$.
\begin{align}
    p(\mu^1_I) & \propto p(\mu_I^0) \exp\left[ -\beta_I \int_0^1 d\lambda  \left( \pdv{V_{II}}{\lambda}  + \pdv{\Omega_e}{\lambda}\Big|_{T_e, \mu_e} \right)  \right] \nonumber \\
        & \propto p(\mu_I^0) \exp\left[ -\beta_I ( V^1_{II} - V^0_{II}  + \Omega_e^1 - \Omega_e^0 )  \right] \nonumber \\
        & \propto \exp\left[ -\beta_I ( K_{II}  + V_{II} + \Omega_e)  \right],
\end{align}
where in the last step we use the kinetic energy of the original state, which hasn't changed. Upon summing over ion particle number, and normalizing, we arrive at the full partition function for the two-temperature electron-ion system,
\begin{align}
\label{eq:BM_partition_function}
    \Xi = \sum_{N_I} \int \prod^{N_I}_I \frac{ d^3\r_I d^3\p_I}{(2\pi\hbar)^{3 N_I}} e^{-\beta_I ( H_{I} - \psi_I + \Omega_{e}    )},
\end{align}
which is related to the total free energy via 
\begin{align}
\Omega = - T_I \ln \Xi.     
\end{align}
We note the resulting partition function is the same as a typical subsystem under the influence of an external force or pressure\cite{Kardar_2007}, but in this case acting individually on each ion. The final answer differs very slightly from Eq.~(4.30) of \cite{BoerckerMore1986} in the way in which electron number is averaged over.

\section{$N$-Temperature Generalization }
Assuming we have a hierarchy of massive particles, each of which has it's own temperature. Then each level acts with the Born-Oppenheimer approximation assumed for all lower levels. Allowing for a quantum-mechanical description of each layer, we have 
\begin{align}
        \Xi_i &= \sum_{N_i} \Tr e^{-\beta_i ( H_{i} - \psi_i + \sum^N_{j>i} H_{i,j}  + \Omega_{i-1}  )} \\
        \Omega_i &= -\frac{1}{\beta_i} \Xi_i
\end{align}
The two-temperature expectation value of an operator is defined through the partition function as,
\begin{align}
    \expv{\mathcal{O}} &= \frac{1}{\Xi} \sum_{N_I,N_e} \frac{\int d^3\r_i d^3\p_i}{(2\pi\hbar)^{3 N_I}} {\rm Tr } \big[ e^{-\beta_I ( H_{I} + \psi_I ) - \beta_e(H_{e} + \psi_e + H_{eI})} \nonumber \\
    & \mathcal{O} e^{-(\beta_I - \beta_e)\Omega_e}\big].
\end{align}
Then the $N-$temperature limit follows immediately
\begin{align}
        \expv{\mathcal{O}} &= \frac{1}{\Xi} \sum_{N_{\{i\}} }\Tr  \big[ \prod_i^N \exp\big[-\beta_i ( H_{i} + \psi_i  + V^{\rm ext}_{i}( \{ r_{j>i} \} ) ) - \nonumber  \\
        &  (\beta_{i+1}-\beta_i)\Omega_i[\{n_{j<i} \}] \big] 
        \mathcal{O} \big],
\end{align}
where the trace is now over every species.

\section{Alternative Ornstein-Zernike Formulation}
\label{app:other_QOZ}
The traditional form\cite{HansenMcdonald} of the Ornstein-Zernike does not involve a matrix inversion and is an implicit equation whose symmetry is not obvious. However, it has use some use, particular when combined with the hypernetted-chain.  
We start with the homogeneous Fourier space limit of Eq.~\eqref{eq:functional_χ_inverse} and combine it directly with Eq.~\eqref{eq:Dyson}, 
\begin{align}
    \chi_{ij}(k) {{\chi^0}^{-1}}_{jj}(k) = \delta_{ij} + \sum_{k} {\chi^{-1}}_{ik}(k) U_{kj}(k).
\end{align}
Where we used the fact that the ideal response function is always species-diagonal. Using the response function in terms of pair correlation functions, Eq.~\eqref{eq:χij_a}-\eqref{eq:χij_d}, and assuming ionic classical response ${\chi^0}_{II}(k) = - \beta_I n_I$, gives the set of equations 
\begin{subequations}
    \begin{align}
        \tilde{h}_{II}(k)   =& -\beta_I U_{II}(k)  - \beta_I\sum_{k}  n_k \tilde{h}_{Ik}(k)  U_{kI}(k) \label{eq:ion-ion_OZ_traditional} \\
        -\frac{\beta_e n_e}{{\chi^0}_{ee}(k)} \tilde{h}_{Ie}(k)   =&  -\beta_e U_{Ie}(k)  - \beta_e \sum_{k} n_k  \tilde{h}_{Ik}(k)  U_{ke}(k) \label{eq:i-e_OZ_traditional} \\
        \tilde{h}_{eI}(k)   =& -\beta_e U_{eI}(k)  - \sum_{k}  n_k \beta_{k}  \tilde{h}_{ek}(k) U_{kI}(k) \nonumber \\
        &+ n_e (\beta_e - \beta_I)\Delta \tilde{h}_{ee}(k)\tilde{U}_{eI}(k)\\
        -\frac{\beta_e n_e}{{\chi^0}_{ee}(k)} \tilde{h}_{ee}(k)   =& n_e^{-1} \left( 1 + \beta_e n_e {\chi^0}^{-1}_{ee}(k)\right) -  \beta_e  U_{ee}(k) \nonumber\\
        & - \sum_{k} \beta_k n_k \tilde{h}_{ek}(k)  U_{ke}(k)\nonumber\\
        & + (\beta_e - \beta_i) \Delta \tilde{h}_{ee}(k) n_e (\tilde{U}_{ee} - {\chi^0}_{ee}^{-1}(k)).
    \end{align}
\end{subequations}
This form of the OZ equations, as opposed to Eq.~\eqref{eq:final_OZ_form1} has useful representations of the $II$ and $Ie$ components for deriving the HNC equations, but with the symmetry of the correlation functions, for example $eI$ and $Ie$, highly obscured.

\end{document}